\documentclass[iop,apj,tighten]{emulateapj}
\usepackage{amsmath,amstext}
\usepackage[breaklinks,colorlinks,citecolor=blue,linkcolor=magenta]{hyperref}
\usepackage[all]{hypcap} %Links go to figures. This breaks deluxetables; use \capstartfalse \capstarttrue around deluxetables to fix it.

\shorttitle{Ejecta Velocities in Tycho's SNR}
\shortauthors{Sato and Hughes}

\begin{document}

\newcommand{\chandra}{{\it Chandra}}
\newcommand{\tycho}{Tycho's SNR}
\newcommand{\eohone}{E0102$-$72}
\newcommand{\kms}{km s$^{-1}$}

\title{Direct Ejecta Velocity Measurements of Tycho's Supernova Remnant}

\author{Toshiki \textsc{Sato}\altaffilmark{1,2,3}
    and John P. \textsc{Hughes}\altaffilmark{3}
   }

\newcommand{\myemail}{toshiki@astro.isas.jaxa.jp}

\altaffiltext{1}{Department of Physics, Tokyo Metropolitan University, 1-1 Minami-Osawa, Hachioji, Tokyo 192-0397}
\altaffiltext{2}{Department of High Energy Astrophysics, Institute of Space and Astronautical Science (ISAS),Japan Aerospace Exploration Agency (JAXA), 3-1-1 Yoshinodai, Sagamihara, 229-8510, Japan; \myemail}
\altaffiltext{3}{Department of Physics and Astronomy, Rutgers University, 136 Frelinghuysen Road, Piscataway, NJ. 08854-8019, USA; jph@physics.rutgers.edu}

\begin{abstract}
We present the first direct ejecta velocity measurements of Tycho's
supernova remnant (SNR). \chandra's high angular resolution images
reveal a patchy structure of radial velocities in the ejecta that can
be separated into distinct redshifted, blueshifted, and low velocity
ejecta clumps or blobs.  The typical velocities of the redshifted and
blueshifted blobs are $\lesssim 7,800$ \kms\ and $\lesssim 5,000$
\kms, respectively. The highest velocity blobs are located near the
center, while the low velocity ones appear near the edge as expected
for a generally spherical expansion. Systematic uncertainty on the
velocity measurements from gain calibration was assessed by carrying
out joint fits of individual blobs with both the ACIS-I and ACIS-S
detectors.   We 
determine the three-dimensional kinematics of the Si- and Fe-rich
clumps in the southeastern quadrant and show that these knots form a
distinct, compact, and kinematically-connected structure, possibly
even a chain of knots strung along the remnant's edge.  By examining
the viewing geometries we conclude that the knots in the southeastern
region are unlikely to be responsible for the high velocity Ca II
absorption features seen in the light echo spectrum of SN~1572, the
originating event for \tycho.
\end{abstract}


\maketitle

%%%%%%%%%%%%%%%
%   Section 1
%%%%%%%%%%%%%%%

\section{Introduction}
The X-ray emission from remnants of Type Ia supernovae (SNe) holds
important clues about the nature of these explosions, which are used
as standardizeable candles to determine the expansion history of the
Universe \citep{riess+1998,perlmutter+1999}.  Additionally there is
increasing evidence that the high-speed shock waves driven by these
explosions accelerate cosmic rays to PeV energies \citep[see][for the
  specific case of Tycho's supernova remnant]{slane+2014}.  To
investigate these two important scientific questions, measurements of
key dynamical quantities, such as ejecta bulk velocity flows,
turbulence, and ion temperatures, are critical, yet such measurements
are extremely challenging to make with current instrumentation.

Tycho's supernova remnant (SNR), recorded by Tycho Brahe in 1572 and
studied by him for more than a year, is known to be the result of a
type Ia supernova (SNIa) based, circumstantially, on the light curve
from Tycho's observations
\citep*{1945ApJ...102..309B,2004Natur.431.1069R} and, definitively, on
the light-echo spectrum obtained with modern instrumentation
\citep{2008Natur.456..617K}.  As the prototypical Galactic example of
a SNIa explosion, Tycho's SNR has been well studied for insights into
the SNIa explosion mechanism.  \cite{2006ApJ...645.1373B} made a
detailed comparison between the ejecta X-ray emission properties of
Tycho's SNR and several SNIa explosion models.  They concluded that
the X-ray morphology and integrated spectrum was well reproduced by a
one-dimensional delayed detonation model with compositionally
stratified ejecta, expanding into a uniform ambient medium density
with density $\rho \sim 2 \times 10^{-24}$ g cm$^{-3}$. Some
collisionless electron heating at the reverse shock ($\beta \equiv
T_e/T_{\rm ion} \sim 0.03$) was necessary to explain the {\it
  XMM-Newton} and {\it Chandra} observations.  In this model, the mean
velocity of the shocked ejecta was estimated to be $\sim$2000 \kms.

The expansion velocities of the forward shock and shocked ejecta have
also been studied through proper motion measurements.
\cite{2000ApJ...545L..53H} made the first accurate X-ray expansion
rate measurement by comparing the brightness profiles from two
observations by the {\it ROSAT} high resolution imager taken in 1990
and 1995.  This indicated expansion rates of
0.22$^{\prime\prime}$--0.44$^{\prime\prime}$ yr$^{-1}$ at the outer
rim of Tycho, where the range represents the variation in expansion
rate from the peak of the ejecta emission to the remnant's edge.
\cite{2010ApJ...709.1387K} used {\it Chandra} observations to measure
the expansion rates of both the forward-shock and the ejecta.  They
found the proper motion of the reverse-shocked ejecta to be
0.21--0.31$^{\prime\prime}$ yr$^{-1}$, consistent with the earlier
{\it ROSAT} work.  Converting these rates into shock velocities
requires knowledge of the remnant's distance which remains uncertain
with a spread of published values mostly between 2 kpc and 4 kpc
\citep[see][for a review of distance determinations to Tycho's
  SNR]{2010ApJ...725..894H}.  For reference, an angular expansion rate
of 0.26$^{\prime\prime}$ yr$^{-1}$ corresponds to a velocity of
$\sim$3700 \kms\ for a distance of 3 kpc.

Spectral measurements have also revealed evidence for significant
ejecta expansion velocities.  Using data from the {\it Suzaku}
satellite, \cite{2009ApJ...693L..61F} and \cite{2010ApJ...725..894H}
found broadened X-ray line spectra from the remnant's interior, which
both studies interpreted as being due to the Doppler shifting of lines
from the approaching and receding hemispheres of the SNR.  They
required expansion velocities of the Si, S and Ar ejecta to be
4700$\pm$100 \kms, somewhat larger than the inferred velocity of the
Fe ejecta (4000$\pm$300 \kms).  The ejecta velocities measured by
these two different methods (proper motion and line broadening) are
broadly consistent and higher than the predicted velocity in
\cite{2006ApJ...645.1373B}.  However, direct measurements of ejecta
velocities in Tycho's SNR have not been done yet.

In this paper, we aim to directly measure the velocities of the
shocked ejecta with {\it Chandra}.  From high angular resolution X-ray
imaging over the years (e.g., with {\it Einstein}, {\it ROSAT} and
{\it Chandra}), clumpy ejecta structures have been clearly noted in
Tycho's SNR
\citep*[e.g.,][]{1983ApJ...266..287S,1995ApJ...441..680V,2002ApJ...581.1101H}.
It is plausible to suspect that such clumps could have different
velocities along the line of sight as a result of, for example, being
located on either the approaching or receding side of the remnant.  If
so it should be possible to separate these with a sufficiently good
combination of X-ray imaging and spectroscopy.  

This paper is organized as follows.  The next section discusses the
observations used and the data reduction procedures applied to the
data.  In \S 3 we present our imaging and spectroscopic analysis of the
data and results on ejecta velocities in Tycho's SNR. Section 4 places
our results in the broader context and the final section concludes.
An Appendix presents an additional validation test of our ejecta
velocity measurements with {\it Chandra}. Throughout this article
uncertainties are quoted at the 90\% confidence level, unless
explicitly stated otherwise.

%%%%%%%%%%%%%%%
%   Section 2
%%%%%%%%%%%%%%%

\section{Observation and Data Reduction}
\subsection{Chandra ACIS-I and ACIS-S Data Sets}
\label{sec:acis}

The {\it Chandra} Advanced CCD Imaging Spectrometer Imaging-array
\citep[ACIS-I,][]{garmire+92,bautz+98} observed Tycho's SNR in April
2009 (PI: Hughes) for a effective exposure of 734.1 ksec.  The
observation was carried out using nine ObsIds as summarized in Table
\ref{tab:chandradata}.  We reprocessed all the level-1 event data,
applying standard data reduction procedures using tasks from version
4.7 of the {\it Chandra} Interactive Analysis of Observations
(CIAO\footnote{Available at {\tt http://cxc.harvard.edu/ciao/}})
package with calibration data from the version 4.6.1 CALDB.  For
spectral extraction, we used {\tt specextract} and made weighted
response files using this script. { The default aspect solution for
  each ObdID was used; previous work has shown that the relative
  registration for the ObsIDs of this Tycho data set is good
  \citep{2011ApJ...728L..28E}.} Unless otherwise stated, background
was taken from the exterior detector area beyond a radius of 4.75
arcmin centered on the remnant. Fits were done using XSPEC 12.8.2 {  and AtomDB 2.0.2}.

\begin{table}[h]
\caption{Log of \chandra\ Observations Used in this Study}
\begin{center}
\begin{tabular}{cccccc}
\hline
         &  	  &	Date 				& Exposure 	& Roll Angle 	\\
Detector & ObsID  &   (YYYY/MM/DD) 		& time (ks)	& (deg)			\\ \hline
ACIS-S  & 115 		& 	2000/09/20		& 	48.9	& 171.0			\\ \hline
ACIS-I  & 10093 		& 	2009/04/13	&	118.4	& 29.2			\\
  "      & 10094 		& 	2009/04/18	&	90.0	& 29.2			\\
  "      & 10095 		& 	2009/04/23	&	173.4	& 29.2			\\
  "      & 10096 		& 	2009/04/27	&	105.7	& 29.2			\\
  "      & 10097 		& 	2009/04/11	&	107.4	& 26.3			\\
  "      & 10902 		& 	2009/04/15	&	39.5	& 29.2			\\
  "      & 10903 		& 	2009/04/17	&	23.9	& 29.2			\\
  "      & 10904 		& 	2009/04/13	& 	34.7	& 29.2			\\
  "      & 10906 		& 	2009/05/03	& 	41.1	& 37.2			\\ \hline
ACIS-I   & sum		&	$\ldots$		&	734.1	&	$\ldots$	\\ \hline
\end{tabular}
\label{tab:chandradata}
\end{center}
\end{table}

\begin{table}[h]
\caption{Log of {\it Suzaku} Observations Used in this Study}
\begin{center}
\begin{tabular}{ccccc}
\hline
     & 	       & Date 		& Exposure (ks)  & 		\\
Name & ObsID   & (YYYY/MM/DD) 	& [XIS0+3]	 & SCI\tablenotemark{a}	\\ \hline
Tycho's SNR & 500024010		&	2006/06/27	&	202.2	 &	off	\\
   "        & 503085020		&	2008/08/11	&	205.7	 &	on	\\ \hline
\eohone\    & 100044030		&	2006/02/02	&	42.6	 &	off 	\\
   "        & 103001030		&	2008/08/12	&	22.6	 &	on 	\\ \hline
\end{tabular}
\tablenotetext{1}{Spaced-row Charge Injection mode (see text)}
\label{tab:suzakudata}
\end{center}
\end{table}

In addition to the frontside-illuminated CCD chips on ACIS-I,
\chandra\ carries a spectroscopic array (ACIS-S) with a
backside-illuminated chip that can be used for imaging.  An
observation of Tycho's SNR using ACIS-S was carried out early in the
mission for an effective exposure time of 48.9 ks (see Table
\ref{tab:chandradata}).  The effective area and spectral resolution of
the ACIS-I and ACIS-S detectors are quite different; additionally the
two detectors allow us to sample two independent sets of readout
electronics for the spectra of individual features in the remnant.
Therefore we utilized both detectors as a powerful cross-check of our
spectral results and to establish the level of systematic error in
derived velocities.  Data reduction and analysis techniques for the
ACIS-S data were the same as for ACIS-I.

\subsection{Suzaku XIS}
\label{sec:suzakuXIS}

For an additional comparison with the {\it Chandra} data, we also
analyzed data from the X-ray Imaging Spectrometers
\citep[XIS,][]{koyama+07} onboard {\it Suzaku}.  The {\it Suzaku} XIS
observed Tycho's SNR twice as summarized in Table
\ref{tab:suzakudata}.  The primary data reduction was performed
following the standard procedures recommended by the instrument team
as implemented in the {\tt aepipeline} task (using
HEASOFT\footnote{Available at {\tt http://heasarc.nasa.gov/lheasoft/}}
version 6.16). Ancillary response files (arf) and redistribution
matrix files (rmf) were generated using {\tt xissimarfgen} and {\tt
  xisrmfgen}, respectively.  For calculating the XIS effective area,
we assumed the \chandra\ image in the 1.6--2 keV band as the input sky
map.  For spectral analysis, we used only the XIS data of the
front-illuminated CCDs (XIS0 and 3).  Background data were taken from
the nearby source-free sky area and subtracted from the source
spectrum.

\begin{figure*}[t]
 \begin{center}
\includegraphics[width=17.0cm]{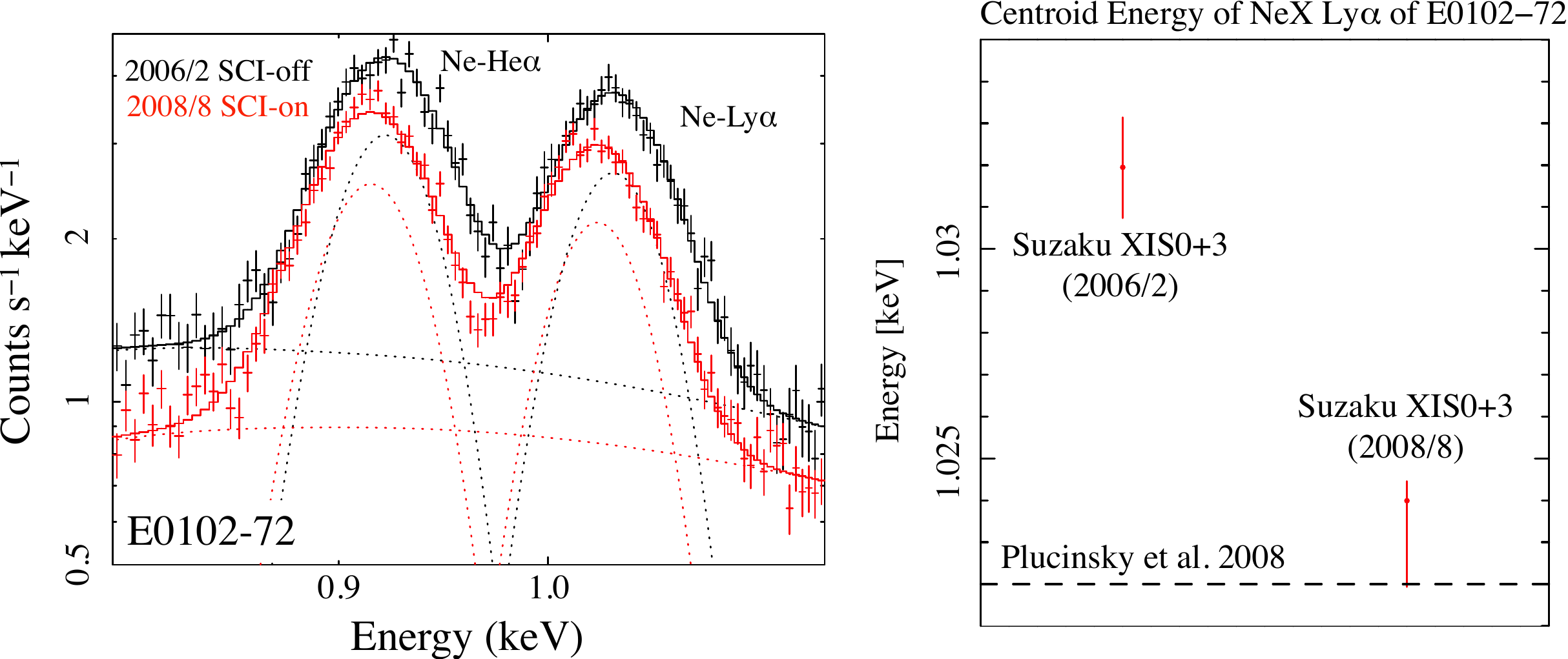}
 \end{center}
\caption{Left: {\it Suzaku} spectra of \eohone\ in the vicinity of the
  Ne X Ly$\alpha$ line from observations taken in 2006 and 2008.
  Right: Comparison of the centroid energy of the Ne X Ly$\alpha$ line
  between 2006 and 2008. The dashed horizontal line shows the expected
  Ne X Ly$\alpha$ line energy of \eohone\ from high spectral resolution
  observations \citep{2016arXiv160703069P}.}
\label{fig:E0102}
\end{figure*}

From October 2006, the XIS observed using the Spaced-row Charge
Injection mode \citep[SCI:][]{2009PASJ...61S...9U}.  The two
observations of Tycho's SNR were therefore observed in each of these
different modes. We found a discrepancy in the fitted line
centroid energy between these two modes.  Figure \ref{fig:E0102} (left
panel) shows the XIS spectra of the calibration source SNR \eohone\ in
these epochs over the energy band that contains the Ne IX He$\alpha$
and Ne X Ly$\alpha$ lines. Fits were done using individual Gaussian
models for the two line features plus a powerlaw continuum.  The Ne X
Ly$\alpha$ line centroid in the 2006 observation is inconsistent by
$\sim$10 eV (Figure \ref{fig:E0102} - right panel) with the value
determined by \cite{2016arXiv160703069P}, who use high spectral
resolution grating instruments to characterize \eohone's 0.3--2.5 keV
band emission and establish this source as an effective calibration
standard.  The centroid energy from our analysis of the 2008 data is,
however, consistent with \cite{2016arXiv160703069P}.

%%%%%%%%%%%%%%%
%   Section 3
%%%%%%%%%%%%%%%

\section{Data Analysis and Results}
\subsection{Radial Profile}
\label{sec:RP}

\begin{figure*}[t]
 \begin{center}
  \includegraphics[width=17cm]{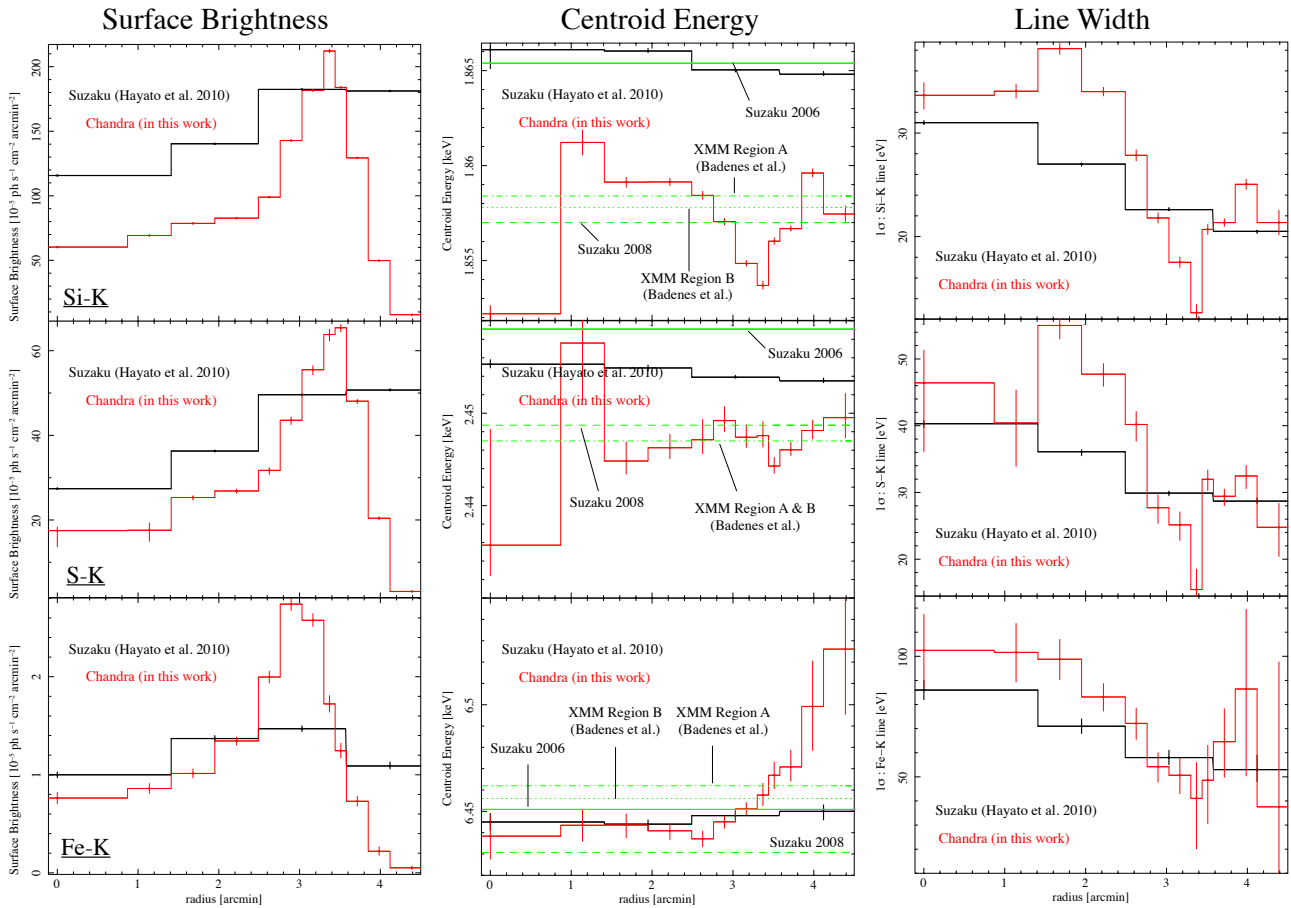}
 \end{center}
\caption{Radial profile of surface brightness, centroid energy and
  line width. Top, middle and bottom show the { Si-He$\alpha$,
    S-He$\alpha$ and Fe-K$\alpha$} lines, respectively. We divided the
  whole SNR into 12 ``Sky" regions. Black curves show the results from
  {\it Suzaku} \citep{2010ApJ...725..894H}, while the red curves show
  the \chandra\ results from this work. The regions from the innermost
  to the outermost are referred to as Sky1--Sky12.  { The regions
    were all centered on the geometric center of the remnant: 00$^{\rm
      h} $25$^{\rm m}$19$^{\rm s}$,
    64$^\circ$08$^\prime$10$^{\prime\prime}$ (J2000). The outer radii
    of the Sky1--Sky12 regions were 0.87$^{\prime}$, 1.41$^{\prime}$,
    1.95$^{\prime}$, 2.49$^{\prime}$, 2.76$^{\prime}$,
    3.03$^{\prime}$, 3.3$^{\prime}$, 3.44$^{\prime}$, 3.58$^{\prime}$,
    3.85$^{\prime}$, 4.12$^{\prime}$ and 4.66$^{\prime}$,
    respectively. The Sky1 to Sky2 regions were complete
    circles/annuli. For the remaining regions (Sky3 to Sky12), we
    excluded the southeast portion of the remnant corresponding to
    position angles between 60$^{\circ}$ and 150$^{\circ}$.}
  Uncertainties are shown at the 90 \% confidence level. Solid
  (dashed) green lines show the best-fit values of centroid energy
  from the entire SNR from 2006 (2008) {\it Suzaku}
  observation. Dash-dotted and dotted green lines show the {\it
    XMM-Newton} results from regions A and B, respectively from
  \citet{2006ApJ...645.1373B}.}
\label{fig:rp}
\end{figure*}

In this section, we check the consistency of the radial
profiles\footnote{As in \cite{2010ApJ...725..894H} we exclude the
  southeastern portion of \tycho\ from our radial profiles, because of
  the failure of a simple shell geometry to describe the images there
  as shown by \cite{2005ApJ...634..376W}}
of line properties from our \chandra\ analysis with past results and,
because of \chandra's sharper point-spread-function (PSF), we also
investigate finer radial dependencies.  Figure \ref{fig:rp} shows the
radial profiles of the surface brightness, centroid energies and line
widths of the K$\alpha$ line blends of Si, S and Fe from {\it Chandra}
(red curves). The black curves show the {\it Suzaku} profiles
\citep{2010ApJ...725..894H}. Four radial bins were used in that work.
In our analysis, we divided essentially the same area into 12 radial
bins (referred to later as Sky1 through Sky12).  \chandra\ spectral
fits followed the same procedure as used in the {\it Suzaku} analysis:
a set a broadened Gaussian lines and a powerlaw continuum fit over the
energy range 1.7--3.4 keV for the Si+S band and 5.0--7.0 for the Fe
band, which also includes the Cr K line.
{ Specifically the lines we included as Gaussians were
  Si-He$\alpha$, Si-Ly$\alpha$, Si-He$\beta$, Si-He$\gamma$,
  S-He$\alpha$, S-Ly$\alpha$, S-He$\beta$ and
  Ar-He$\alpha$(+S-He$\beta$) in the Si+S band and Cr-K$\alpha$ and
  Fe-K$\alpha$ in the Fe band.  We treated the He-like 
  complexes as single broadened Gaussians with their centroid energies
  and normalizations as free parameters. The line widths of the
  prominent He$\alpha$ blends of Si, S, Ar were fitted freely. For the
  other blends (e.g., He$\beta$, He$\gamma$ and Ly$\alpha$), we fixed
  the line width to the He$\alpha$ value. The intensity, centroid, and
  width of the Fe-K$\alpha$ line were free parameters; the width of
  the Cr-K$\alpha$ line was fixed to that of the Fe line. We note that
  the energy centroid of the Fe K$\alpha$ line in Tycho is $\sim$6.4
  keV and corrresponds to a mean ionization state of Fe XVII
  \citep*{2014ApJ...780..136Y}; over the entire remnant it can be
  described well by a single broadened Gaussian line.}

The radial bins in the \chandra\ profiles are all fully independent,
unlike the {\it Suzaku} profiles where the broader PSF of the {\it
  Suzaku} X-ray telescope (with a half-power diameter of
$\sim$2$^\prime$) causes significant amounts of flux to mix from one
bin to the others. Consequently, we obtain much sharper surface
brightness profiles than {\it Suzaku}.  We found that the Si-He$\alpha$ and the
S-He$\alpha$ lines peak in intensity at a radius of
$\sim$3.3$^{\prime}$--3.5$^{\prime}$, consistent with {\it XMM-Newton}
\citep*{2001A&A...365L.218D}.  We note that the S-He$\alpha$ line appears to
peak at a slightly higher radius than the Si-He$\alpha$ line. %, which may be
%related to the earlier heating of the outer layers of ejecta as
%discussed in \cite{2015ApJ...805..142L}.
For Fe-K$\alpha$, we found an intensity peak at a radius of
$\sim$3.0$^{\prime}$, which is also consistent with past results
\citep*[e.g.,][]{2005ApJ...634..376W,2014ApJ...780..136Y}.

The energy centroid profiles are shown in the middle columns of Figure
\ref{fig:rp}.  Again the \chandra\ profiles are shown in red and the
published {\it Suzaku} ones are in black.  The green lines show other
values integrated over the remnant from the literature
\citep[XMM:][]{2006ApJ...645.1373B} or our own analysis of the 2006
and 2008 {\it Suzaku} data.  For the Si and S line centroid energies,
the new \chandra\ profiles are inconsistent with the {\it Suzaku} data
taken in 2006, which are $\sim$10 eV too high.  This is also the data
set with inconsistent line centroids for the calibration target
\eohone, so we are justified in ignoring it for this consistency
check.  Since the \chandra\ profiles are consistent with all the other
data sets, we are confident that the energy scale of the
\chandra\ observation is accurate.

In principle, for a perfectly uniform emitting shell, the energy
centroid profiles should be flat.  In fact, however, the
\chandra\ profiles show significant radial structure. One notable
feature is the statistically significant jump in centroid energy from
the innermost bin to the second one.  For the Si-He$\alpha$ line, this jump is
$\sim$9 eV, which corresponds to a difference in line-of-sight
velocity of $\sim$1450 \kms.  This value is about 30\% of the
expansion speed of the Si-rich shell and it can be explained if we
assume that there is an factor of approximately two difference in the
intrinsic intensity of the approaching and receding hemispheres in
this radial bin. We propose that the patchy nature of the remnant's
emission is the source of the structures seen in the Si and S line
centroid profiles. We consider this in further detail in
\S~\ref{sec:largescale} below.  We also consider the increasing Fe line
centroid energy beyond the peak emission below
(\S~\ref{sec:fe-increase}).

For the line width profile, we found a gradual decrease from the
center toward the edge. This feature was also seen (albeit at lower
resolution) by {\it Suzaku}
\citep*{2009ApJ...693L..61F,2010ApJ...725..894H}; these authors
interpreted the variation of the line width radial profile as the
signature of an expanding shell of ejecta.  An important new feature
of the \chandra\ profiles is the clear minimum in the line width at a
radius of $\sim$3.4$^{\prime}$ (the Sky8 region). This is also where
the line intensity peaks. We identify this as the region where the
ejecta are moving most closely to the plane of the sky and therefore
show little to no Doppler shift.

\subsection{Expansion Velocity}
\label{sec:expvel}

\begin{table*}[t]
\caption{Best-fit Parameters of the Double Gaussian Model from the Central Regions\tablenotemark{a} Using {\it Chandra}}
\begin{center}
\small
\begin{tabular}{cccccccc}
\hline
 	& Width\tablenotemark{b} & $E_{\rm red}$ & $E_{\rm blue}$ & 2$\delta E$ 	& $v_\perp$\tablenotemark{c} & $v_{\rm exp}$\tablenotemark{c} (\kms)	 & $v_{\rm exp}$ (\kms)	\\
Lines	& (eV)	& (keV)	       & (keV) 	& (eV) 	&	(\kms)	& (this work)	&	(Hayato et al.~2010)				\\\hline
Si-He$\alpha$ 						& 		12.6	&		1.821$\pm$0.003				&		1.880$^{+0.002}_{-0.003}$	&		59$\pm$4			&		4780$\pm$320			&		5010$\pm$340			&		4730$^{+30}_{-20}$			\\
S-He$\alpha$						& 	15			&		2.393$^{+0.006}_{-0.009}$	&		2.478$^{+0.004}_{-0.006}$	&		87$^{+7}_{-11}$		&		5240$^{+430}_{-680}$	&		5490$^{+450}_{-710}$	&		4660$\pm$50		\\
Fe-K$\alpha$						& 		41		&		6.36$\pm$0.02				&		6.53$\pm$0.02				&		170$\pm$30			&		4000$\pm$700			&		4200$\pm$800			&		4000$\pm$300				\\\hline
\multicolumn{8}{c}{Using only high surface brightness pixels} \\
Si-He$\alpha$ 		& 		12.6	&		1.822$\pm$0.003				&		1.884$^{+0.002}_{-0.003}$	&		62$\pm$4			&		5020$\pm$320			&		5260$\pm$340			&		---			\\
S-He$\alpha$		& 	15			&		2.389$\pm$0.007				&		2.478$^{+0.004}_{-0.005}$	&		89$^{+8}_{-9}$		&		5490$^{+490}_{-550}$	&		5750$^{+510}_{-580}$	&		---		\\\hline
\multicolumn{8}{c}{Using only low surface brightness pixels} \\
Si-He$\alpha$		& 		12.6	&		1.821$^{+0.004}_{-0.005}$	&		1.880$^{+0.003}_{-0.004}$	&		59$^{+4}_{-5}$		&		4780$^{+320}_{-410}$	&		5010$^{+340}_{-430}$	&		---			\\
S-He$\alpha$		& 	15			&		2.395$^{+0.008}_{-0.009}$	&		2.479$^{+0.007}_{-0.008}$	&		84$^{+11}_{-12}$	&		5170$^{+680}_{-740}$	&		5410$^{+710}_{-770}$	&		---		\\\hline
\end{tabular}
\tablenotetext{1}{From within the central 1.41$^\prime$ radius (Sky1+Sky2 regions).}
\tablenotetext{2}{Fixed at the appropriate minimum values from the radial profiles (for region Sky 8).}
\tablenotetext{3}{ Velocities are based on the centroid shifts of the Gaussian models.}
\label{tab:exp}
\end{center}
\end{table*}

\begin{figure}[h]
 \begin{center}
 \includegraphics[width=8cm]{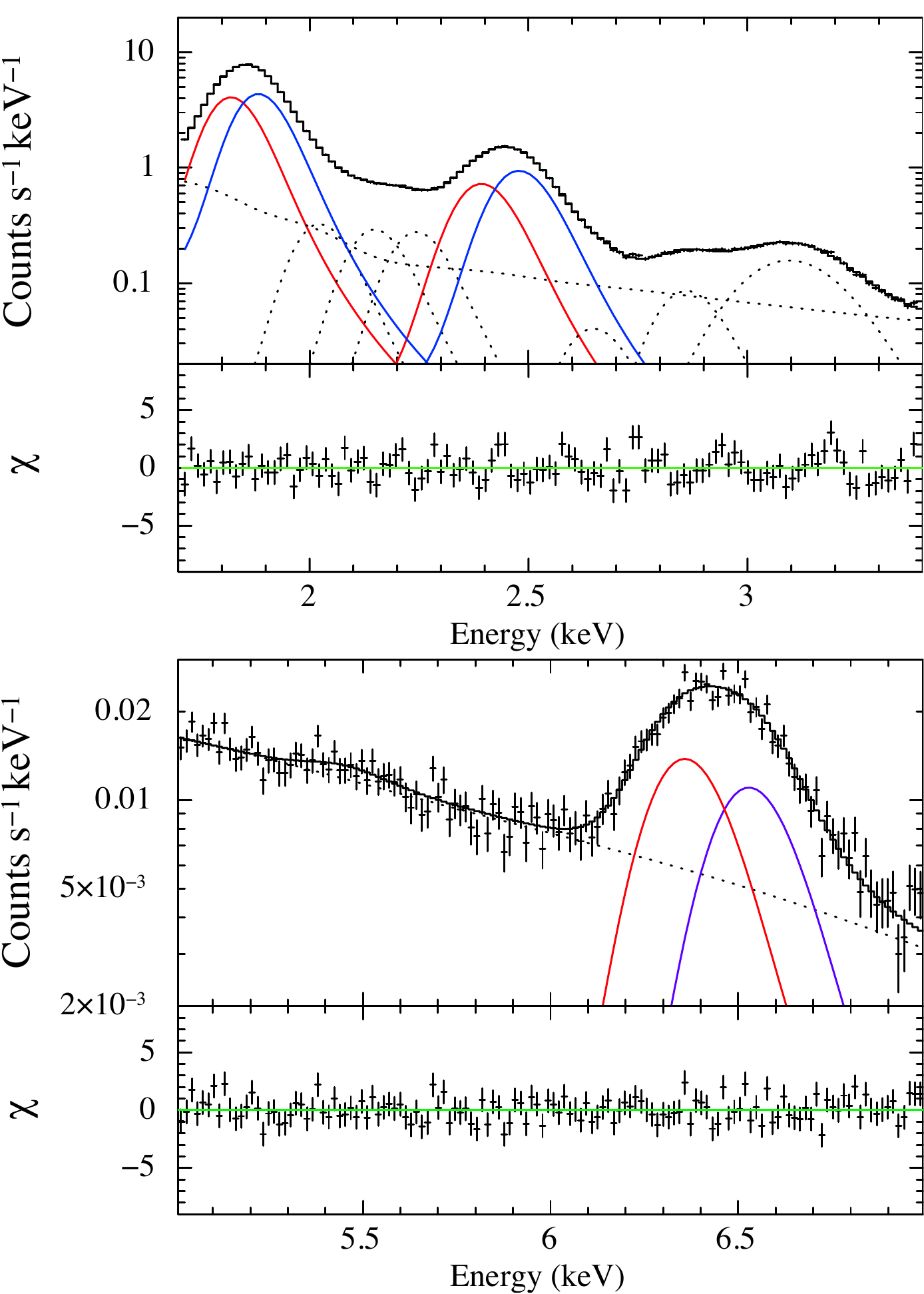}
 \end{center}
\caption{Double Gaussian model fit to the \chandra\ spectrum from the
  center of \tycho\ (Sky1+Sky2 regions) in the Si+S band (top panel)
  and the Fe-K$\alpha$ band (bottom panel). The weak bump at $\sim$5.6
  keV in the bottom panel is Cr K$\alpha$ line emission. The continuum
  emission is fitted with a power-law model.}
\label{fig:specdoublegauss}
\end{figure}

Here we estimate the shell expansion velocity from the Si, S, and Fe K
lines using the \chandra\ data from the center of \tycho, extracting
the spectrum from within a radius of 1.41$^\prime$ (Sky1 plus Sky2
regions) to match the previous work with {\it Suzaku}.  We modeled the
line broadening as in previous work \citep{2010ApJ...725..894H} with
two Gaussian lines corresponding to the Doppler-shifted components
from the receding and approaching hemispheres of the expanding shell
of ejecta. { The faint lines (specifically Ly$\alpha$, He$\beta$
  and He$\gamma$) were included in several ways: as single narrow or
  broadened lines, or as narrow lines linked in velocity to the
  corresponding red- or blueshifted He$\alpha$ line.  We also allowed
  the width of each of the double He$\alpha$ Gaussian lines to vary
  within the allowed range from the Sky8 region fits ($^{+0.5}_{-0.6}$
  eV for Si-He$\alpha$ and $\pm$2 eV for S-He$\alpha$). In all cases
  the derived energy differences between the red- and blueshifted
  components for both Si and S were in agreement.  The final
  uncertainty on the energy difference includes the statistical
  uncertainty plus the range from the different spectral models.}
Results of the \chandra\ fits for the three species are given in Table
\ref{tab:exp} { (for now we only consider the top three entries
  that correspond to the integrated spectrum from the central
  region).}  For the Si+S band the best fit yields $\chi^2=146.75$ for
92 degrees of freedom and for the Fe band $\chi^2=137.32$ for 127
degrees of freedom. Figure~\ref{fig:specdoublegauss} shows that the
double Gaussian model is a good fit to the spectra data.
%Uncertainties in the table are quoted at the 90\% confidence level.

For the spectral fits, line widths were fixed at the values determined
from the minima in each radial profile.  The radial velocity
($v_\perp$) was calculated from the average energy of the red- and
blue-shifted components and $\delta E$. To convert $v_\perp$ to the
shell expansion speed we needed to correct for the projection factor
over the central 1.41$^\prime$ of the remnant.  For Si-He$\alpha$ and
S-He$\alpha$, we assumed a spherically symmetric shell extending over
190$^{\prime\prime}$--220$^{\prime\prime}$ and for Fe-K$\alpha$, we
assumed the shell covered the radial range
180$^{\prime\prime}$--200$^{\prime\prime}$. We calculated the
projection factor using the method in section A.1 of
\cite{2010ApJ...725..894H}. We determined projection factors of 0.955
for Si-He$\alpha$ and S-He$\alpha$ and 0.948 for Fe-K$\alpha$.  For
{\it Suzaku}, an additional correction was necessary to account for
the flux spreading due to the broad {\it Suzaku} PSF.  This can be
ignored for the {\it Chandra} analysis.

The shell expansion velocities from \chandra\ are in the range of
$\sim$4200--5500 \kms; all are consistent with the {\it Suzaku}
results.  The large uncertainty on the Fe shell expansion from
\chandra\ alone does not allow us to exclude that it is moving at the
same speed as the other species.  However, by combining all of the Si
and S measurements (both {\it Chandra} and {\it Suzaku}, i.e.,
  using a weighted combination of the values in the last two columns
  of Table~\ref{tab:exp}) we arrive at an expansion velocity for the
Si+S shell of $4724^{+27}_{-19}$ \kms\ that is
significantly greater ($>$4$\sigma$) than the (similarly
  combined) Fe-shell expansion velocity of $4025\pm 280$ \kms.

{ The results quoted above correspond to an
  emission-measure-weighted mean velocity difference for the
  approaching and receding hemispheres. Although this is a practical
  approach from an observational perspective, it may not produce an
  unbiased result.
%This is not the only (or even
%  the best) way to characterize the velocity difference and in fact it
%  may even be a biased estimator.  
  If, for example, there is a correlation between expansion speed and
  emission-measure (say, because higher intensity spots, such as
  compact blobs of ejecta, tend to expand more quickly) then we will
  get a biased velocity difference.  However, thanks to \chandra's
  high angular resolution we can assess this effect.  We divided all
  pixels within the central 1.41$^{\prime}$ region by surface
  brightness into either high or low values (splitting at the mean
  level of $\sim$$6\times10^{-4}$ ph s$^{-1}$ cm$^{-2}$ arcmin$^{-2}$)
  and extracted a separate spectrum for each pixel set.  Results for
  the Si and S lines are given in the bottom entries of Table
  \ref{tab:exp} and in all cases (Si and S, high and low surface
  brightness), the velocity differences are statistically in agreement
  with the result from the integrated spectrum. This suggests that in
  general there is not a strong correlation between expansion speed
  and the brightness of features through the center of Tycho's SNR. }

\subsection{Mean Photon Energy Map of Si-He$\alpha$}
\label{sec:meanphotonmap}

\begin{figure*}[t]
 \begin{center}
  \includegraphics[trim= 50 150 50 150,clip,width=8cm]{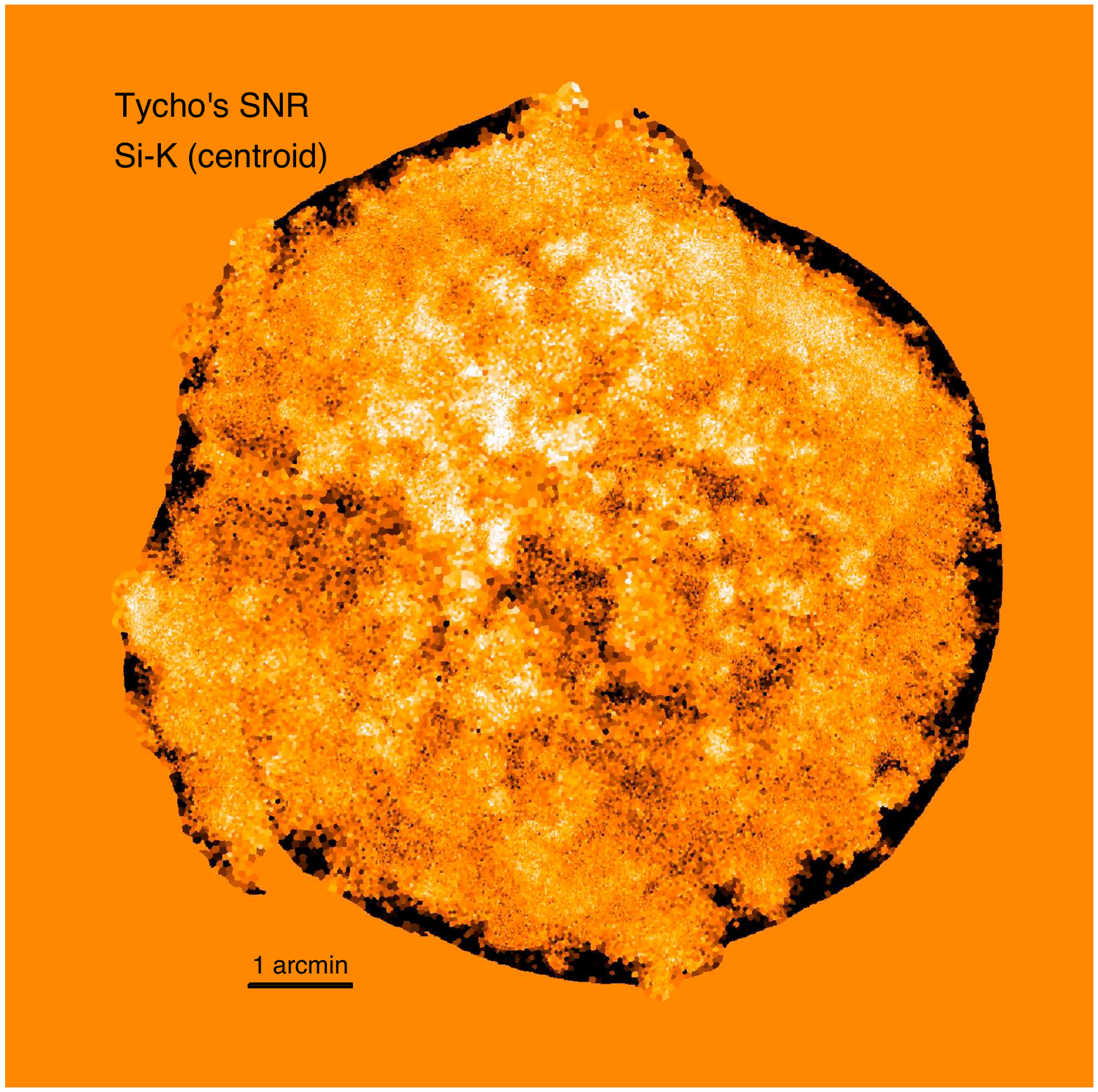}
  \includegraphics[trim= 50 150 50 150,clip,width=8cm]{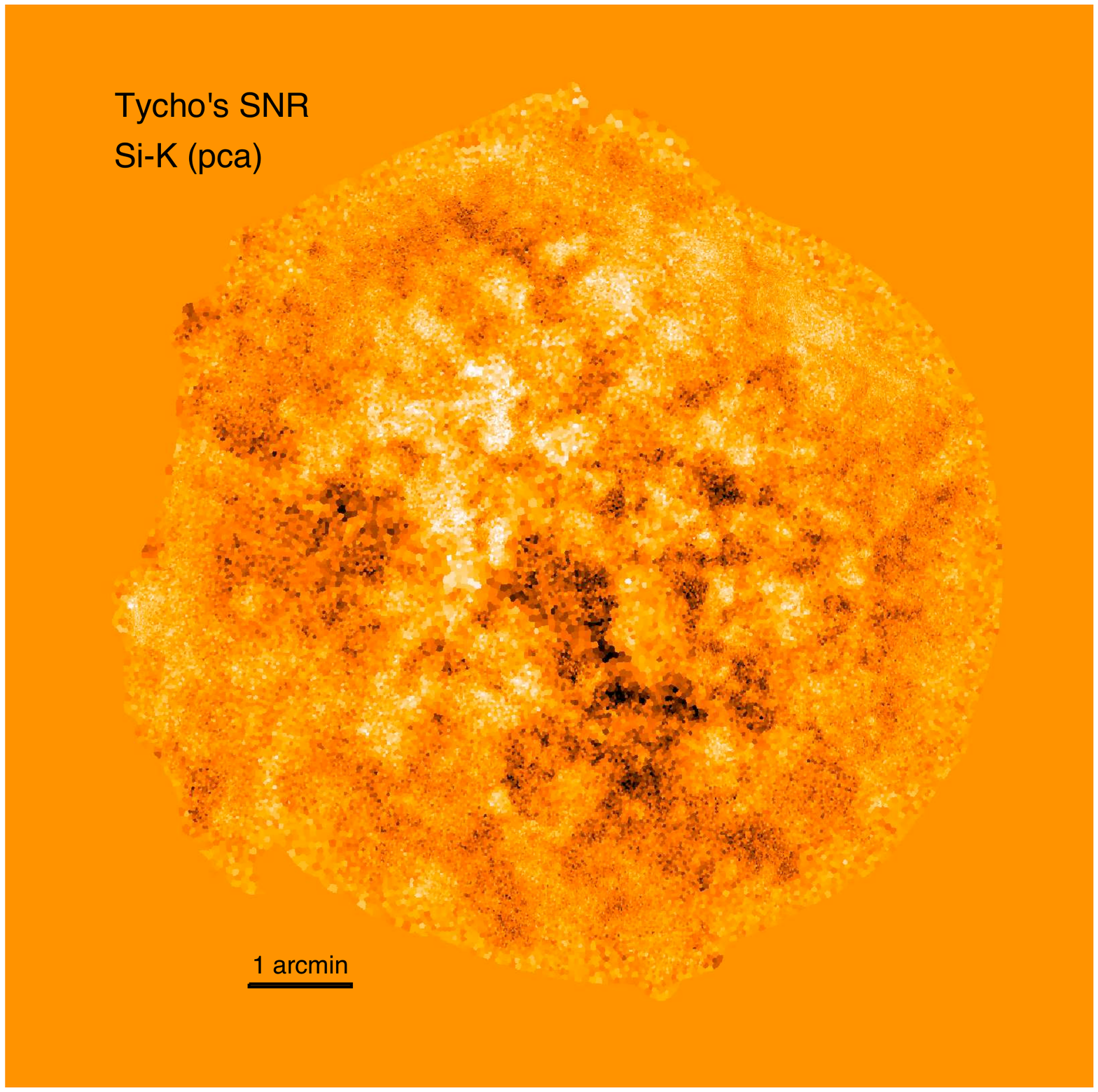}
 \end{center}
\caption{Left: Mean photon energy map in the Si-He$\alpha$ band (1.6--2.1 keV)
  from the deep \chandra\ ACIS-I observation of Tycho's SNR.  Voroni
  Tessellation was used to combine pixels to produce varying-sized
  regions with similar signal-to-noise ratio in each region. We chose
  a S/N of 20 for this image (i.e., approximately 400 detected Si line
  photons). The color scale varies linearly from energy centroid values
  of 1.816 keV (dark) to 1.879 (light).  The dark ring around the edge
  is where nonthermal dominates over thermal emission. Right: Map of
  the Si-He$\alpha$ band data projected onto the Principal Component that
  separates red- and blue-shifted emission (see text).  The color
  scale here has been adjusted to approximately match that in the left
  panel.}
\label{fig:img}
\end{figure*}

Small clumpy structures in the ejecta in Tycho's SNR have been noted
since the {\it Einstein Observatory} High Resolution Imager
observations in the late 1970s \citep*[e.g.,][]{1983ApJ...266..287S}.
We consider here the possibility that these structures or blobs might
have different expansion speeds which would be manifest as differences
in line centroid energy due to the Doppler effect. For this we used the
Si-He$\alpha$ line because of its large statistical signal.

We divided the 1.6--2.1 keV band into 34 energy bins { (each 15 eV
  in width)} and made fluxed images at each energy using the {\tt
  merge\_obs} script in CIAO.  We computed the mean photon energy in
each bin using
\begin{equation*}
E_{\rm mean} =  \frac{\sum_{i} E_i I_i}{\sum_{i} I_i},
\end{equation*}
where $E_i$ and $I_i$ are the photon energy and the intensity at each
energy bin ($i = 1,2,...,34$).  To equalize the signal-to-noise across
the map, we used Voroni Tessellation
\citep*[e.g.,][]{2003MNRAS.342..345C,2006MNRAS.368..497D} to merge
pixels together to reach a uniform signal-to-noise ratio of 20 in each
bin.  Hereafter we refer to these as VT bins.

Figure \ref{fig:img} (left panel) shows the mean photon energy map for
the Si-He$\alpha$ line in Tycho's SNR. The image is dominated by patchy
structures that, in many cases but not all, can be associated with
specific features in the intensity map. There is some striping in the
image that correlates with the readout direction of the chips (toward
the NE and SW), but this effect is clearly 
subdominant to the patchy structure of the remnant. The dark region 
around the rim of the remnant is where nonthermal continuum emission 
dominates; we do not remove the continuum in our map-making procedure 
so these regions are not correct in this map.  The maximum range of mean 
photon energy values is $\gtrsim$60 eV, which corresponds to a range of 
ejecta velocities of $\gtrsim$9,700 km s$^{-1}$. This is quite close to 
twice the Si-He$\alpha$ expansion velocity determined above (see Table 
\ref{tab:exp}).

We have also done a Principal Components Analysis (PCA) of the Si-He$\alpha$
band images following closely the previous application of this method
to Tycho's SNR \citep{2005ApJ...634..376W}. In our application here,
we generated a large number of Si-He$\alpha$ line spectra (one from each VT bin)
and compressed each from 34 spectral bins into 18 by { summing together bins in}
the fainter wings of the line profile.  These spectra were input to the
PCA algorithm \citep{murtaghheck87} to identify new axes in the
 18-dimensional space of the data set that maximized its variance.

As discussed by \cite{2005ApJ...634..376W}, there is no guarantee, in
general, that the Principal Components (PCs) identified by PCA have a
unique astrophysical interpretation.  However, in this case of a
specific emission line, we found that the first three PCs have
spectral templates with a clear and unequivocal interpretation: (1)
line equivalent width, (2) line energy centroid, and (3) line energy
width.  There are as many PCs as spectral bins (18 here) and in the
case of a totally random dataset, each PC would account for $\sim$6\%
of the variance of the full data set.  Here we found that the first
three PCs each account for 17\%, 15\% and 6\% of the variance with the
remaining components each accounting for less than 5\%.  We conclude
therefore that the first two PCs are significant, the third is
marginal, and that the remaining PCs are all insignificant. 

A map of the \chandra\ Si-He$\alpha$ spectral data projected onto the second PC
is shown in Figure \ref{fig:img} (right panel). This map closely
matches the mean centroid map shown in the left panel validating our
interpretation of it as being due to line centroid variations.  The
agreement between the two maps is poor near the outer rim, where
continuum emission causes the centroid calculation to produce spurious
results.

Together these maps indicate that there is enough variation in the
Si-He$\alpha$ line centroids to motivate identifying individual red- and
blue-shifted blobs and measuring their velocities through spectral
analysis.  We turn to this in the next section.

\subsection{Spectral Analysis of Specific Blobs}
\label{sec:blobs}

\begin{figure*}[t]
 \begin{center}
  \includegraphics[trim=0 0 150 0, clip,width=12cm]{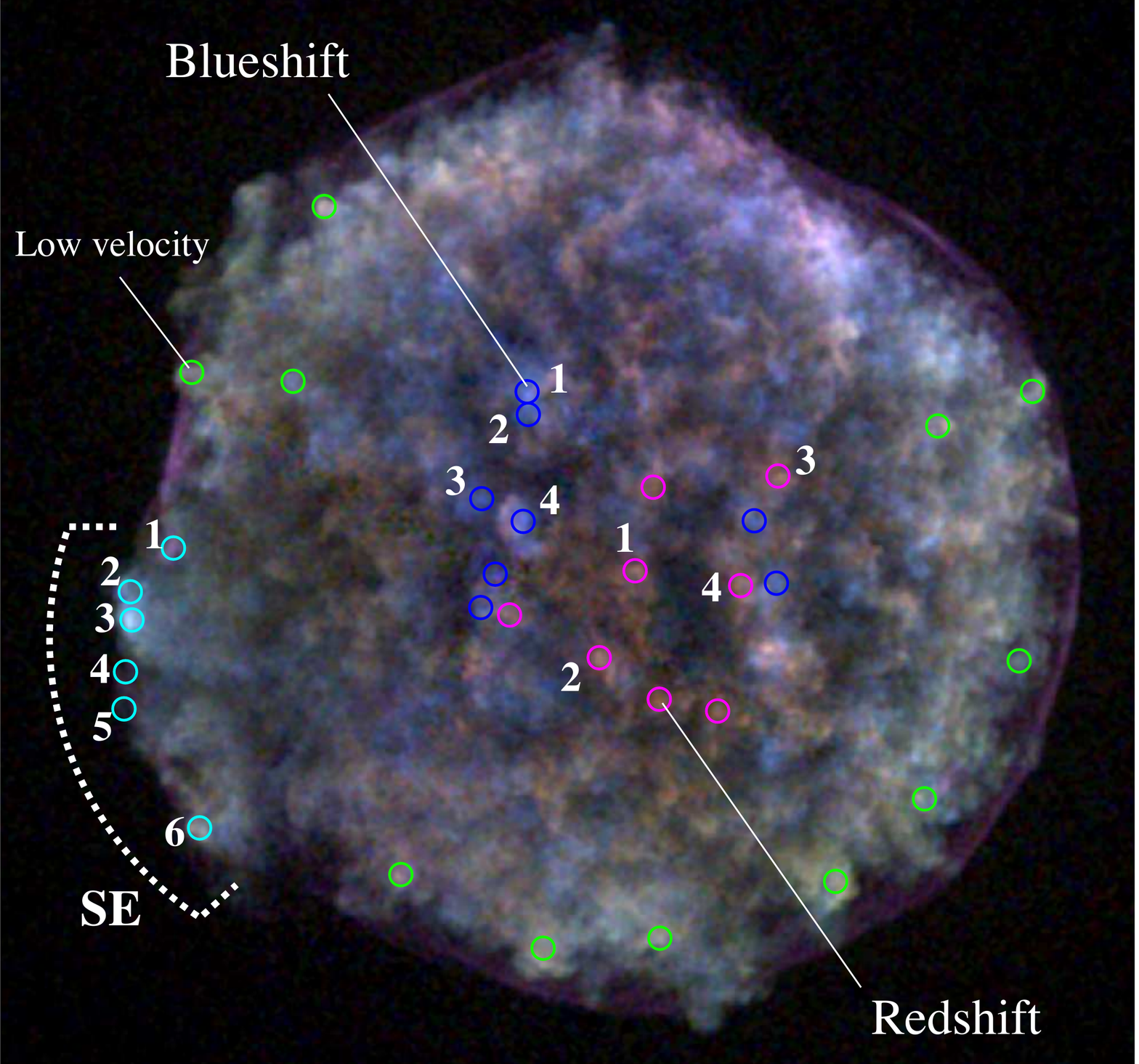}
 \end{center}
\caption{Three-color image of the Si-He$\alpha$ line from the \chandra\ ACIS-I
  observation of Tycho's SNR.  The red, green and blue images come
  from the 1.7666--1.7812 keV, 1.8396--1.8542 keV, and 1.9564--1.971
  keV bands. Magneta, blue, and green circles identify the redshifted,
  blueshifted and low velocity blobs, respectively, used for the
  spectral analysis. Likewise the cyan circles show the knots in the
  southeastern quadrant that we studied.}
\label{fig:three_color}
\end{figure*}

\begin{figure}[h]
 \begin{center}
    \includegraphics[clip,width=8cm]{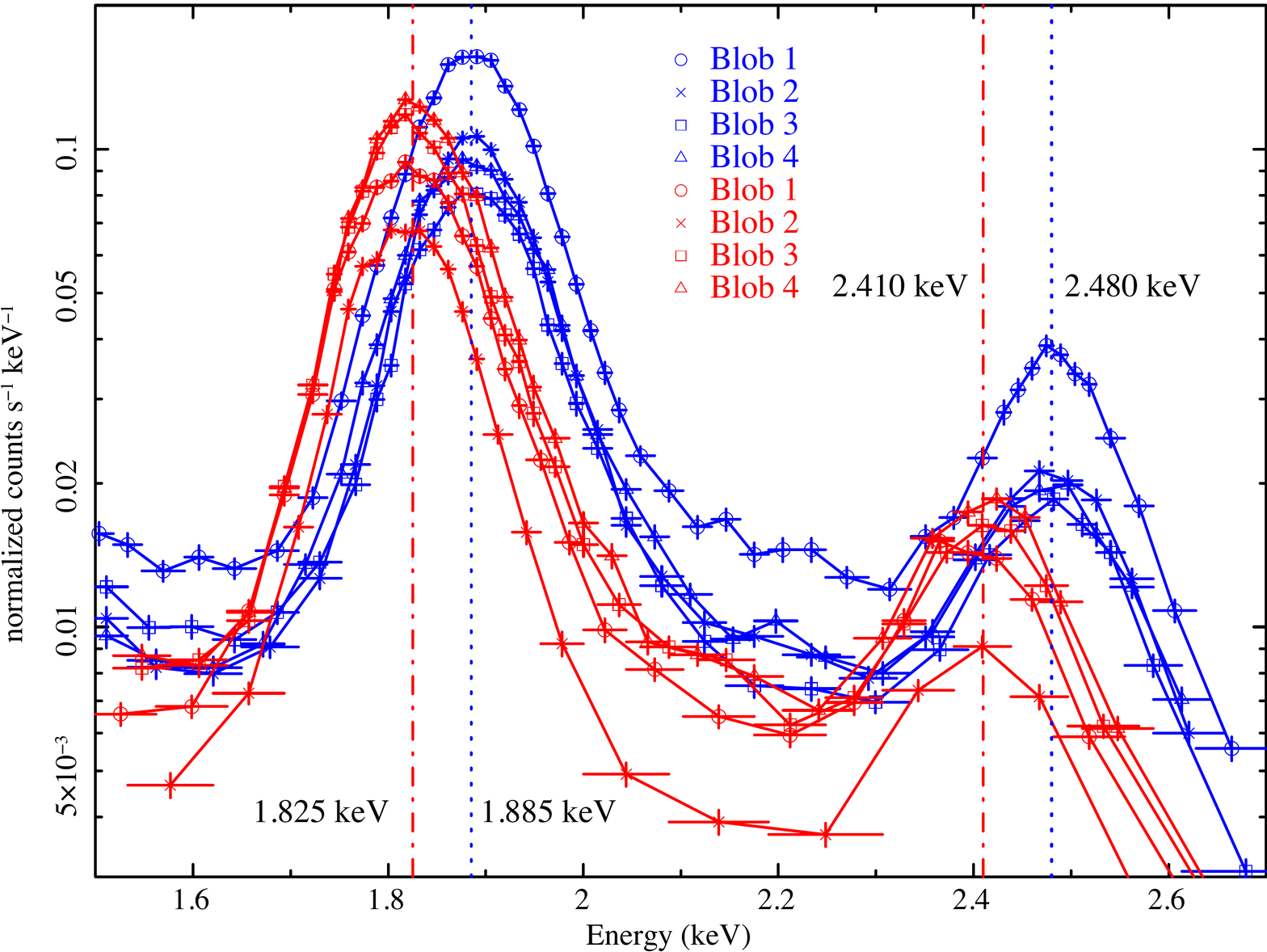}
 \end{center}
\caption{Typical spectra of the red- and blue-shifted blobs. The
  symbol types and numeric labels correspond to the regions from which
  the spectra were extracted shown on Figure~\ref{fig:three_color}.
  Vertical bars show the 1 $\sigma$ uncertainty on intensity;
  horizontal bars just indicate the size of the energy bin.}
\label{fig:img_spec}
\end{figure}

Figure \ref{fig:three_color} shows a color image of Tycho's SNR
constructed from three narrow energy slices of the Si-He$\alpha$ line
as noted in the figure caption.  In this figure, we see two kinds of
Si-He$\alpha$ blobs in the central region: blobs with higher (bluish
color) or lower (reddish color) photon energy. We defined
appropriately sized, { circular} regions { (r = 0.1$^{\prime}$)}
for these blobs and extracted their spectra (shown in Figure
\ref{fig:img_spec}).  There is a clear separation of the centroid
energies between the red- and the blue-shifted blobs. { Rough eyeball
estimates for the Si-He$\alpha$ and the S-He$\alpha$ lines suggest
centroid energy differences of $\sim$60 eV and $\sim$70 eV,
respectively, or Doppler velocity differences of $\sim$9,000 km
s$^{-1}$.}

In addition to the Doppler effect, line centroids are also sensitive
to the shock-heating history of the line-emitting material. In
particular the ionization age (which is the product of the electron
density and the time since the material was shock heated, $n_e t$)
{ can affect the prominence of the H-like Ly$\alpha$ line and, for
  lower values, can also affect the energy centroid of the He-like
  complex}.  Increasing the ionization age tends to increase the mean
charge state which tends to increase the centroid of the K
line. However, over rather wide ranges of ionization ages, the charge
state is dominated by the He-like species (as in the case of \tycho)
and the dependence of line centroid on ionization age is weak.
Furthermore, increases in charge state from He-like to H-like produce
noticeable distortions in the shape of the Si K line (i.e., making it
double peaked) even at CCD spectral resolution.  We see no significant
evidence for { such} line shape distortions in the PCA results {
  (i.e., the significant PCs correspond to the first three moments of
  the line: intensity, centroid, and width, while there are no
  significant PCs corresponding to higher moments, such as skewness or
  bimodality)}, so large ionization age variations are not expected.

Nevertheless, to separate Doppler shifts from any possible ionization
state changes, we carried out detailed spectral analyses using
nonequilibium ionization (NEI) models.  We extracted the spectra from
27 regions in total including the red- and blue-shifted blobs
mentioned above, as well as several low velocity blobs near the edge
of \tycho.  These were fitted with the vnei model (for the NEI thermal
component) and the srcut model (for the nonthermal continuum component)
in XSPEC. Also, we allowed for the model spectra to be broadened using
the gsmooth model since thermal broadening and/or multiple Doppler
components might be present.  For the srcut model, we assumed a
constant radio spectral index of $\alpha = -0.65$
\citep{2006A&A...457.1081K} based on the integrated flux densities at
408 and 1420 MHz and allowed the cutoff frequency and radio intensity
to be free parameters. Absorption due to the intervening column
density of interstellar material is negligible in this band ($>$1.6
keV) so we ignored it for these fits.

\begin{figure}[h]
 \begin{center}
  \includegraphics[trim=0 35 0 5,clip,width=8.5cm]{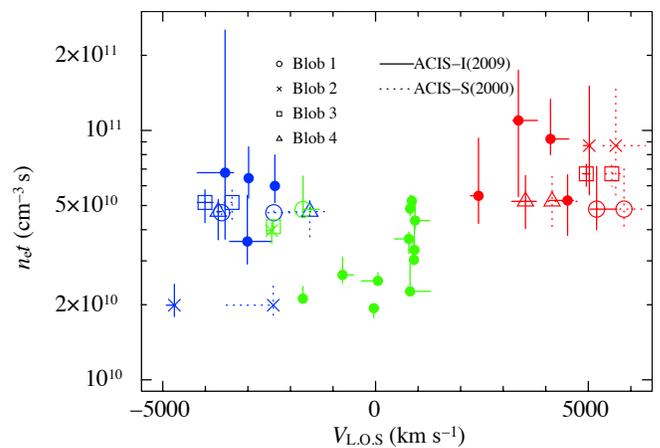}
 \end{center}
\caption{Scatter plot between the line-of-sight velocity and the
  ionization age ($n_e t$) for each blob. Velocities are based on
    the vnei model ``redshift'' parameter. The open symbols,
  identification numbers, and red or blue colors correspond to the 8
  regions in Figure \ref{fig:img_spec}. Solid and dashed error bars
  show results from the ACIS-I and ACIS-S detectors, respectively. The
  filled circles show the results of the other 19 regions in Figure
  \ref{fig:img_spec} and the colors correspond to redshifted (red),
  blueshifted (blue) or low velocity (green) blobs.}
\label{fig:V}
\end{figure}

Figure \ref{fig:V} shows a scatter plot of the best-fit line-of-sight
velocity versus the ionization age for each blob.  The maximum
separation of blob velocity reaches $\sim$9000 \kms\ even taking into
account the variability of the ionization age. From the thermal model,
the ionization ages are in the range of $\sim10^{10}$--$10^{11}$
cm$^{-3}$ s, and electron temperatures are $\sim$0.9--2.7 keV (mean
$kT\sim$1.3 keV).  We also analyzed a number of blobs close to the
edge of the remnant. These blobs have smaller velocities than the
blobs from the interior, but a similar range of ionization timescales.
The pattern of line-of-sight velocities shown in Figure \ref{fig:V} is
consistent with the effect of projection on the line-of-sight
velocities and agrees qualitatively with the results in Figure
\ref{fig:rp}.

{ The detector gain for ACIS is monitored and updated by regular
  observation of the external $^{55}$Fe source onboard
  \chandra. However, since} there is no simultaneous,
independent gain reference for the ACIS-I detector during any specific
observation, we are potentially subject to uncalibrated gain
variations.  In order to assess this effect we extracted matched
spectra of 8 blobs from the ACIS-S detector and fit them using the
same model as above.  In this analysis, we initially conducted a joint
fit between the ACIS-I and the ACIS-S data for each blob. Then we
linked all parameters except for the gsmooth and redshift parameters
and fitted for independent values of the broadening and velocity.

The ACIS-S spectral fitting results are shown in Figure \ref{fig:V}
using the same symbol types as for the ACIS-I results, except now with
dashed error bars.  Table \ref{tab:blob_id} gives the sky locations of
the jointly fitted blobs, their locations on the detector (i.e., CCD
chip and readout node), and the best-fit velocity for the two data
sets. We obtained similar velocities in the two ACIS data sets.  There
is a discrepancy of $\sim$500--2,000 \kms\ in the sense that the
ACIS-S detector tends to yield more redshifted spectra than ACIS-I.
%This might be a gain effect of each detector.
However, even when averaging all of the velocity measurements, we
still see a velocity difference of $>$8,000 \kms\ between the red- and
blue-shifted blobs.  Thus we conclude that our spectral separation of
the red- and the blue-shifted components correspond to intrinsic
velocity differences in Tycho's SNR.  Velocity measurements, however,
carry a systematic uncertainty of $\sim$500--2,000 \kms.  Improvements
of the ACIS gain calibration may help to reduce this systematic error.

\begin{table*}[t]
\caption{Summary of Joint ACIS-I and ACIS-S Spectral Analysis of Red- and Blue-shifted Blobs}\label{tab:blob_id}
\begin{center}
\scriptsize
\begin{tabular}{ccccccccccccc}
\hline
&										&																							& 		\multicolumn{2}{c}{ACIS-I} 			&		ACIS-S 			&			\multicolumn{2}{c}{Sky Background}					&			\multicolumn{2}{c}{Blob Local Background}							&\\
&	id									&		 (R.A., Decl.)																		&		chip 			&		node		&		node 			&	$V_{\rm I}$\tablenotemark{a} [\kms]	&$V_{\rm S}$\tablenotemark{b} [\kms]	&	$V_{\rm I}$\tablenotemark{a} [\kms]	&	$V_{\rm S}$\tablenotemark{b} [\kms]			&\\ \hline
\multicolumn{3}{l}{\bf Blueshifted blobs}& 																							&						&					&								\multicolumn{2}{c}{(Mean: $-3220\pm$970)}				&		\multicolumn{2}{c}{(Mean: $-4310\pm$880)}						&\\
&Blob1									& (00$^{\rm h}$25$^{\rm m}$24$^{\rm s}$.952, 64$^\circ$09$^\prime$33$^{\prime\prime}$.76)	&			2			&			0		&			1			&		$-3616^{+4}_{-88}$		&		$-2390^{+700}_{-140}$	&		$-4880^{+140}_{-30}$	&		$-3440^{+700}_{-140}$			&\\
&Blob2									& (00$^{\rm h}$25$^{\rm m}$24$^{\rm s}$.843, 64$^\circ$09$^\prime$21$^{\prime\prime}$.72)	&			2			&			0		&			1			&		$-4730^{+30}_{-190}$	&		$-2400^{+10}_{-1100}$	&		$-5030^{+170}_{-40}$	&		$-3350^{+30}_{-160}$			&\\
&Blob3									& (00$^{\rm h}$25$^{\rm m}$28$^{\rm s}$.633, 64$^\circ$08$^\prime$37$^{\prime\prime}$.08)	&			2			&			0		&			1, 2		&		$-4000^{+190}_{-180}$	&		$-3370^{+60}_{-190}$	&		$-5040^{+140}_{-50}$	&		$-4910^{+90}_{-900}$			&\\
&Blob4									& (00$^{\rm h}$25$^{\rm m}$25$^{\rm s}$.275, 64$^\circ$08$^\prime$25$^{\prime\prime}$.04)	&			2, 3		&			0, 3	&			1			&		$-3700^{+80}_{-40}$		&		$-1550^{+420}_{-500}$	&		$-5030^{+50}_{-70}$		&		$-2790^{+500}_{-900}$			&\\
&Blob5									& (00$^{\rm h}$25$^{\rm m}$27$^{\rm s}$.538, 64$^\circ$07$^\prime$57$^{\prime\prime}$.05)	&			0, 1, 2		&			0, 3	&			---			&		$-3540^{+200}_{-660}$	&		---						&		---						&		---								&\\
&Blob6									& (00$^{\rm h}$25$^{\rm m}$28$^{\rm s}$.710, 64$^\circ$07$^\prime$39$^{\prime\prime}$.33)	&			0, 1		&			0, 3	&			---			&		$-3010^{+580}_{-420}$	&		---						&		---						&		---								&\\
&Blob7									& (00$^{\rm h}$25$^{\rm m}$06$^{\rm s}$.514, 64$^\circ$08$^\prime$25$^{\prime\prime}$.35)	&			3			&			3		&			---			&		$-2370^{+26}_{-40}$		&		---						&		---						&		---								&\\
&Blob8									& (00$^{\rm h}$25$^{\rm m}$04$^{\rm s}$.715, 64$^\circ$07$^\prime$52$^{\prime\prime}$.27)	&			3			&			3		&			---			&		$-2984^{+4}_{-1}$		&		---						&		---						&		---								&\\
\multicolumn{3}{l}{\bf Redshifted blobs}	& 																						&						&					&								\multicolumn{2}{c}{(Mean: $+4980\pm$740)}				&		\multicolumn{2}{c}{(Mean: $+7230\pm$840)}						&\\
&Blob1									& (00$^{\rm h}$25$^{\rm m}$16$^{\rm s}$.180, 64$^\circ$07$^\prime$58$^{\prime\prime}$.82)	&			3			&			3		&			1			&		$+5200^{+480}_{-150}$	&		$+5840^{+470}_{-100}$	&		$+7780^{+420}_{-220}$	&		$+7580^{+710}_{-310}$			&\\
&Blob2									& (00$^{\rm h}$25$^{\rm m}$14$^{\rm s}$.237, 64$^\circ$06$^\prime$50$^{\prime\prime}$.80)	&			1			&			0		&			1			&		$+5020^{+20}_{-140}$	&		$+5650^{+690}_{-300}$	&		$+7580^{+710}_{-210}$	&		$+7420^{+1870}_{-680}$			&\\
&Blob3									& (00$^{\rm h}$25$^{\rm m}$04$^{\rm s}$.588, 64$^\circ$08$^\prime$49$^{\prime\prime}$.07)	&			3			&			2, 3	&			0			&		$+4950\pm90$			&		$+5550^{+200}_{-150}$	&		$+7580^{+680}_{-180}$	&		$+7610^{+1480}_{-170}$			&\\
&Blob4									& (00$^{\rm h}$25$^{\rm m}$07$^{\rm s}$.629, 64$^\circ$07$^\prime$50$^{\prime\prime}$.99)	&			3			&			3		&			0, 1		&		$+3500^{+260}_{-320}$	&		$+4150^{+680}_{-90}$	&		$+5040^{+140}_{-110}$	&		$+7210\pm330$					&\\
&Blob5									& (00$^{\rm h}$25$^{\rm m}$14$^{\rm s}$.709, 64$^\circ$08$^\prime$43$^{\prime\prime}$.20)	&			0, 1		&			0, 3	&			---			&		$+2420^{+50}_{-190}$	&		---						&		---						&		---								&\\
&Blob6									& (00$^{\rm h}$25$^{\rm m}$26$^{\rm s}$.368, 64$^\circ$07$^\prime$35$^{\prime\prime}$.43)	&			1			&			0		&			---			&		$+3360^{+450}_{-130}$	&		---						&		---						&		---								&\\
&Blob7									& (00$^{\rm h}$25$^{\rm m}$14$^{\rm s}$.237, 64$^\circ$06$^\prime$50$^{\prime\prime}$.80)	&			1, 3		&			0, 3	&			---			&		$+4510^{+140}_{-430}$	&		---						&		---						&		---								&\\
&Blob8									& (00$^{\rm h}$25$^{\rm m}$09$^{\rm s}$.489, 64$^\circ$06$^\prime$44$^{\prime\prime}$.50)	&			3			&			3		&			---			&		$+4110^{+430}_{-30}$	&		---						&		---						&		---								&\\
\multicolumn{3}{l}{\bf Low velocity blobs}	& 																						&						&					&																																								&\\
&Blob1									& (00$^{\rm h}$25$^{\rm m}$41$^{\rm s}$.454, 64$^\circ$11$^\prime$11$^{\prime\prime}$.91)	&			2			&			0, 1	&			---			&		$+850^{+30}_{-40}$		&		---						&		---						&		---								&\\
&Blob2									& (00$^{\rm h}$25$^{\rm m}$43$^{\rm s}$.965, 64$^\circ$09$^\prime$39$^{\prime\prime}$.19)	&			0, 2		&			0, 3	&			---			&		$-1700^{+390}_{-10}$	&		---						&		---						&		---								&\\
&Blob3									& (00$^{\rm h}$25$^{\rm m}$52$^{\rm s}$.184, 64$^\circ$09$^\prime$43$^{\prime\prime}$.82)	&			0, 2		&			0, 3	&			---			&		$-40^{+120}_{-50}$		&		---						&		---						&		---								&\\
&Blob4									& (00$^{\rm h}$25$^{\rm m}$35$^{\rm s}$.171, 64$^\circ$05$^\prime$17$^{\prime\prime}$.71)	&			1			&			1		&			---			&		$+810^{+481}_{-3}$		&		---						&		---						&		---								&\\
&Blob5									& (00$^{\rm h}$25$^{\rm m}$23$^{\rm s}$.641, 64$^\circ$04$^\prime$38$^{\prime\prime}$.78)	&			1			&			1		&			---			&		$+811^{+117}_{-1}$		&		---						&		---						&		---								&\\
&Blob6									& (00$^{\rm h}$25$^{\rm m}$14$^{\rm s}$.187, 64$^\circ$04$^\prime$44$^{\prime\prime}$.08)	&			1			&			0, 1	&			---			&		$+930^{+350}_{-30}$		&		---						&		---						&		---								&\\
&Blob7									& (00$^{\rm h}$24$^{\rm m}$59$^{\rm s}$.954, 64$^\circ$05$^\prime$14$^{\prime\prime}$.10)	&			1			&			0, 1	&			---			&		$+780^{+20}_{-340}$		&		---						&		---						&		---								&\\
&Blob8									& (00$^{\rm h}$24$^{\rm m}$52$^{\rm s}$.735, 64$^\circ$05$^\prime$57$^{\prime\prime}$.71)	&			1, 3		&			0, 3	&			---			&		$-780^{+270}_{-60}$		&		---						&		---						&		---								&\\
&Blob9									& (00$^{\rm h}$24$^{\rm m}$45$^{\rm s}$.049, 64$^\circ$07$^\prime$10$^{\prime\prime}$.81)	&			3			&			2, 3	&			---			&		$-1710^{+50}_{-20}$		&		---						&		---						&		---								&\\
&Blob10									& (00$^{\rm h}$25$^{\rm m}$51$^{\rm s}$.594, 64$^\circ$09$^\prime$15$^{\prime\prime}$.48)	&			3			&			2		&			---			&		$+910^{+10}_{-130}$		&		---						&		---						&		---								&\\
&Blob11									& (00$^{\rm h}$24$^{\rm m}$43$^{\rm s}$.913, 64$^\circ$09$^\prime$33$^{\prime\prime}$.67)	&			3			&			2		&			---			&		$+60^{+30}_{-390}$		&		---						&		---						&		---								&\\ \hline
\end{tabular}
\end{center}
\tablenotetext{1}{Line-of-sight velocity using the ACIS-I detector. 
Velocities are based on the vnei model ``redshift'' parameter.}
\tablenotetext{2}{Line-of-sight velocity using the ACIS-S detector. 
Velocities are based on the vnei model ``redshift'' parameter.}
\end{table*}

\begin{figure}[t]
 \begin{center}
  \includegraphics[clip,width=8cm]{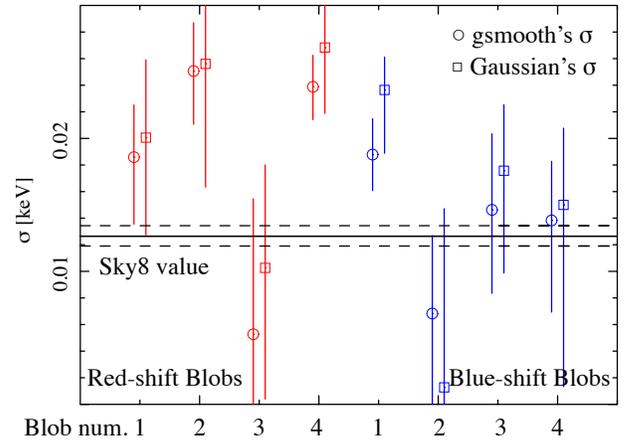}
 \end{center}
\caption{Best-fit Si-He$\alpha$ line widths for 8 individual blobs using local
  regions near each blob for background. Circle (box) symbols show the
  results of fits using the gsmooth (Gaussian lines) model.  Solid and
  dashed lines show the best-fit value and 90\% confidence level
  uncertainty from the Sky8 region.  }
\label{fig:sigma}
\end{figure}

Finally we consider the possibility of contamination of a blob's
spectrum from material in the extraction region at a different
velocity (from, for example, the other side of the shell).  Such
contamination would tend to reduce a blob's observed velocity compared
to its actual velocity. To assess this effect, we extracted local
background spectra from regions near each blob (the fits presented
above used spectra of blank-sky regions from beyond the remnant's
edge) and carried out the spectral fits with the new background
spectra.  In Table \ref{tab:blob_id}, we summarize the fit results
under the columns labeled ``Blob Local Background.''  Not surprisingly
we found best-fit velocities higher by $\sim$1,000-2,000 \kms\ than
with the traditional blank-sky background. Additionally best-fit line
widths were smaller.  In the case of the blank-sky background, line
widths were in the range of $\sim$20--40 eV, while with the local blob
background, line widths were typically a factor of two lower and
generally consistent with the minimum line widths obtained from the
Sky8 region (see Figure \ref{fig:sigma} for the comparison).  These
results suggest that there is some contamination from different
velocity components in the blob spectra and that the actual velocities
of the blobs could be as high as in the last two columns (``Blob
  Local Background'') of Table \ref{tab:blob_id}.  However, the
highly structured nature of the X-ray emission on arcsecond scales
makes a {\it precise} determination of the amount of contaminating
material in any individual blob's spectrum difficult to do in
practice.  However, as an ensemble, it is plausible to conclude that
fits using local blob background spectra provide reasonable
upperbounds on the velocities of the red- and blue-shifted blobs of
$\lesssim 7,800$ km s$^{-1}$ and $\lesssim 5,000$ km s$^{-1}$,
respectively.

\subsection{Large Scale Distribution of Apparent Ejecta Velocity}
\label{sec:largescale}

Here we focus on the large scale velocity structure of \tycho.
Inspection of the mean photon energy map (Figure \ref{fig:img} shows
an obvious asymmetry in the distribution of red- and blue-shifted
blobs between the northern and southern sides.  This asymmetry is also
clearly visible in the ACIS-S data (image not shown).

\begin{figure*}[t]
 \begin{center}
  \includegraphics[trim=0 0 0 250,clip,width=18cm]{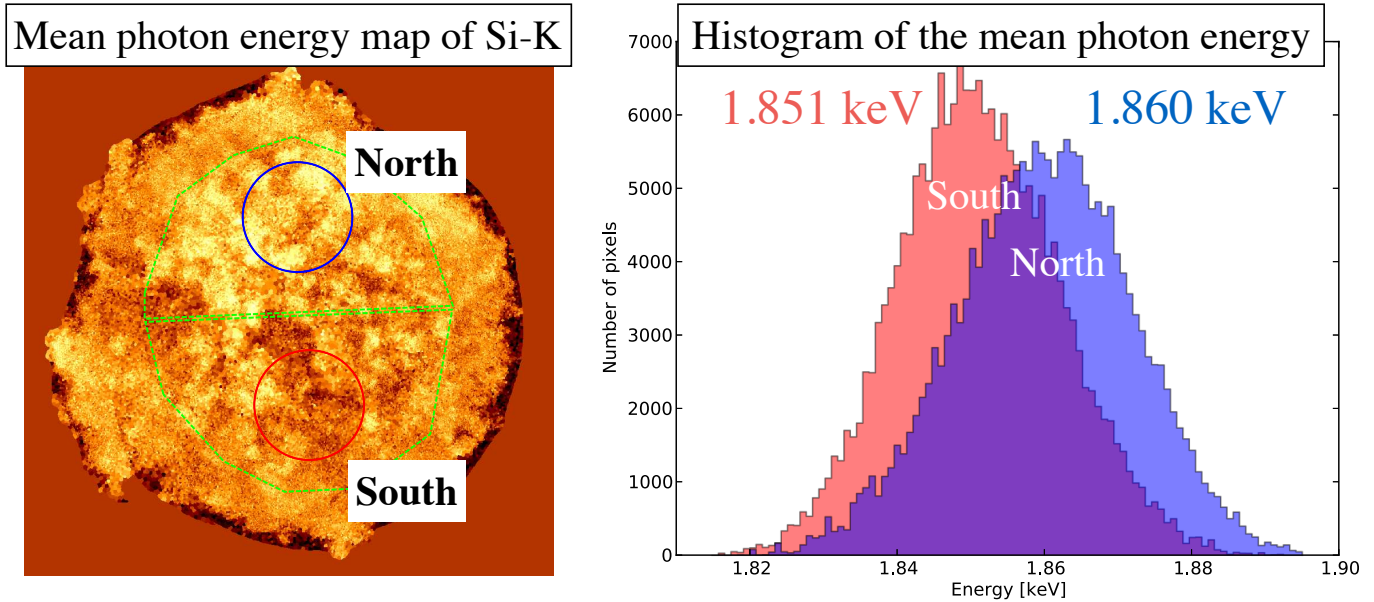}
 \end{center}
\caption{Left: definition of the northern and southern side regions
  with the mean photon energy map. Two green polygon regions are used
  for making the histogram. Two circles are used for the spectral
  analysis. Right: the histogram of the number of pixels in the two
  green semicircular regions as shown in the left figure.}
\label{fig:AE}
\end{figure*}

When the remnant is divided into northern and southern parts (using
the two green semicircular regions shown in Figure \ref{fig:AE},
left), we find different mean energies: $\sim$1.860 kev from the north
and $\sim$1.851 keV from the south (Figure \ref{fig:AE} right).  The
mean energy of the Si-He$\alpha$ line at the edge of the remnant (in the Sky8
region where the line width is minimum), determined using the same
method, is $\sim$1.856 keV, which is approximately halfway between the
energies of the two halves just determined. Taking this energy as the
``rest frame'' of \tycho\ we find that the bulk of the northern and
southern halves appear to be moving along the line-of-sight at
$\sim$$\pm$700 \kms\ with respect to this frame.

\begin{table}[h]
\caption{Line Centroid Energies in the Northern and Southern Regions of Tycho's SNR with {\it Chandra} \& {\it Suzaku}}
\begin{center}
\scriptsize
\begin{tabular}{l|ccc}
\hline
		 							&		ACIS-I						&		 	ACIS-S 							&		 	{\it Suzaku} 	 					\\ \hline
{\underline {\bf Si-He$\alpha$}} 	&									&		 		 							&		 				 	 					\\
North (keV)							& 	1.8634$\pm$0.0003				&		1.8613$\pm$0.0005					&		1.8571$^{+0.0003}_{-0.0002}$			\\
South (keV)							& 	1.8493$\pm$0.0003				&		1.8490$\pm$0.0007					&		1.8545$\pm$0.0003						\\
$\delta E$ (eV)						& 	14.1$\pm$0.4					&		12.3$\pm$0.9						&		2.7$^{+0.4}_{-0.3}$						\\
$\delta V$ (km s$^{-1}$)\tablenotemark{a}			& 	2280$\pm$60						&		1990$\pm$150						&		440$^{+60}_{-50}$						\\ \hline
{\underline {\bf S--He$\alpha$}}		&									&											&												\\
North (keV)							& 	2.4494$\pm$0.0007				&		2.4481$^{+0.0035}_{-0.0027}$		&		2.4506$^{+0.0017}_{-0.0010}$			\\
South (keV)							& 	2.4318$^{+0.0017}_{-0.0014}$	&		2.4282$^{+0.0047}_{-0.0030}$		&		2.4441$^{+0.0016}_{-0.0014}$			\\
$\delta E$ (eV)						& 	17.6$^{+1.8}_{-1.6}$			&		19.9$^{+5.9}_{-4.0}$				&		6.5$^{+2.3}_{-1.7}$						\\
$\delta V$ (km s$^{-1}$)\tablenotemark{a}			& 	2160$^{+220}_{-200}$			&		2450$^{+730}_{-490}$				&		800$^{+280}_{-210}$						\\ \hline
{\underline {\bf Fe-K$\alpha$}}		&									&											&												\\
North (keV)							& 	6.452$^{+0.006}_{-0.007}$		&		6.435$^{+0.014}_{-0.012}$			&		6.430$\pm$0.005							\\
South (keV)							& 	6.413$\pm$0.006					&		6.415$\pm$0.014						&		6.420$^{+0.006}_{-0.007}$				\\
$\delta E$ (eV)						& 	39$^{+8}_{-9}$					&		19$^{+20}_{-19}$					&		10$^{+8}_{-9}$							\\
$\delta V$ (km s$^{-1}$)\tablenotemark{a}			& 	1820$^{+370}_{-420}$			&		890$^{+930}_{-890}$					&		690$^{+550}_{-620}$						\\ \hline
\end{tabular}
\label{tab:AE}
\end{center}
\tablenotetext{1}{Velocities are based on the centroid shifts of the Gaussian models.}
\end{table}

To examine this issue in more detail we extracted ACIS-I, ACIS-S, and
{\it Suzaku} XIS spectra from the two circular regions shown in Figure
\ref{fig:AE} (the northern region is centered at [R.A., Decl.] =
  [00$^{\rm h}$25$^{\rm m}$17$^{\rm s}$.519,
    64$^\circ$10$^\prime$06$^{\prime\prime}$.53], the southern one is
  at [R.A., Decl.] = [00$^{\rm h}$25$^{\rm m}$15$^{\rm s}$.581,
    64$^\circ$06$^\prime$41$^{\prime\prime}$.07], and both regions are
  1 arcmin in radius). Fits were done using the same Gaussian model
as in \S~\ref{sec:RP}. Results are given in Table \ref{tab:AE}.  We
found a strong tendency for the northern region to have higher
centroid energies than the southern region. Of particular note is the
excellent numerical consistency between the ACIS-I and ACIS-S
detectors, which demonstrates that this effect is not an observational
artifact.  For the Si-He$\alpha$ and S-He$\alpha$ line, the difference
of centroid energies corresponds to a line-of-sight velocity
difference of $\sim$2000 km s$^{-1}$.  The result for the {\it Suzaku}
XIS is not as large due to the smoothing induced by {\it Suzaku}'s
broad PSF.

Before jumping to the conclusion that this velocity difference implies
a kinematic asymmetry in the original SN explosion, we first must
explore the possibility that the velocity difference is due to the
patchy nature of the ejecta shell.  Recall that we interpret the
fluctuations in the mean line centroid in the radial profiles
(Figure~\ref{fig:rp}) in this way.  We can estimate the average
velocities in the northern and southern regions assuming different
relative amounts of emission from the approaching and receding
hemispheres using this simple algebraic function
\begin{equation*}
 \langle v_{N,S}\rangle  = \frac{v_{\rm red}I_{{\rm red},N,S}+v_{\rm blue}I_{{\rm blue},N,S}}{I_{{\rm red},N,S}+I_{{\rm blue},N,S}},
\end{equation*}
where the labels ``red'' and ``blue'' indicate the red and blueshifted
components, the labels ``N'' and ``S'' refer to the northern and
southern sides, and $I$ is the line intensity from each of the four
relevant locations.  Then $\langle v_{N,S}\rangle$ are the average
speeds in each region. Next we define the relative intensity ratio
$\chi_{N,S} \equiv I_{{\rm blue},N,S}/I_{{\rm red},N,S}$ allowing us
to define the difference of the average velocities between the
northern and southern regions as
\begin{equation*}
 \langle v_{N}\rangle - \langle v_{S}\rangle = v \left(\frac{1-\chi_N}{1+\chi_N} - \frac{1-\chi_S}{1+\chi_S} \right),
\end{equation*}
where we have made the simplest assumption that the north and south
regions have the identical expansion speed, $v$, along the line of
sight. We choose the northern and southern circular regions to be
offset symmetrically from the center so this assumption is
reasonable. Because of the offset, the radial speed in each region is
less than the shell expansion speed (Table~\ref{tab:exp}), by a
projection factor.  For a Si shell radius of $\sim$3.4$^{\prime}$ the
offset location of the circular regions ($\sim$1.7$^{\prime}$) yields
a projection factor to the line-of-sight of $\cos \, 30^\circ =
0.866$, which yields a projected speed of $v\sim 4000$ \kms.  The
observed north-south velocity difference is $\sim$ $-2000$ \kms, so
the value in the parentheses of the above equation is $-0.5$.  This
value can be accommodated by a range of front-back intensity ratios
for the north and south in the (physically plausible) range $\chi_N
=1/3$ and $\chi_S=0$ to $\chi_N =3$ and $\chi_S=1$. The most modest
intensity ratio differences are $\chi_N =5/3$ and $\chi_S =3/5$, less
than a factor of two for each side.

Thus a biased {\it intensity} distribution for a uniformly expanding
shell of ejecta can account for the observed north to south {\it
  velocity} difference in \tycho. And the biased intensity
distribution is not necessarily the result of an asymmetric explosion,
since local variations in the ambient medium density can result in
significant local intensity differences in the ejecta.  A higher
ambient medium density is likely why the ejecta emission is so much
brighter in the northwest quadrant of \tycho\ \citep[see,
  e.g.,][]{2010ApJ...709.1387K}.  To explain the velocity difference,
we suggest that this enhancement extends over the front (blueshifted)
part of the shell but not over the back (redshifted) part.  This would
require that the mean ejecta density be $\lesssim$ $\sqrt{3}$ higher in
the front than the back, which is plausible given the estimated
ambient density enhancement of a factor of $\sim$2 in the northwest
\citep{2010ApJ...709.1387K}.

\subsection{Velocities of Southeastern Knots}
\label{sec:VSK}

The southeastern (SE) quadrant of \tycho\ is morphologically and
compositionally different from the rest of the remnant. For example,
the bright knots\footnote{
We refer to these features in the SE, historically identified by
composition and brightness, as ``knots'' to distinguish them from the
``blobs'' introduced in this article that are features identified by
radial velocity.}
in the SE are located at a radius of $\sim$4.2$^{\prime}$, which is
$\sim$20\%-30\% further out from the center than the peak
Si-He$\alpha$ and Fe-K$\alpha$ line intensity over the rest of the
remnant.  In addition, the SE knots show strong differences in
relative Si to Fe abundances \citep[e.g.,][]{1995ApJ...441..680V,
  2001A&A...365L.218D,2005ApJ...634..376W}.  Here we study the
kinematic properties of compact knots in this region, focusing on the
six knots identified in Figure \ref{fig:three_color} (cyan circles).

\begin{figure*}[h]
 \begin{center}
  \includegraphics[trim=0 0 0 150,clip,width=18cm]{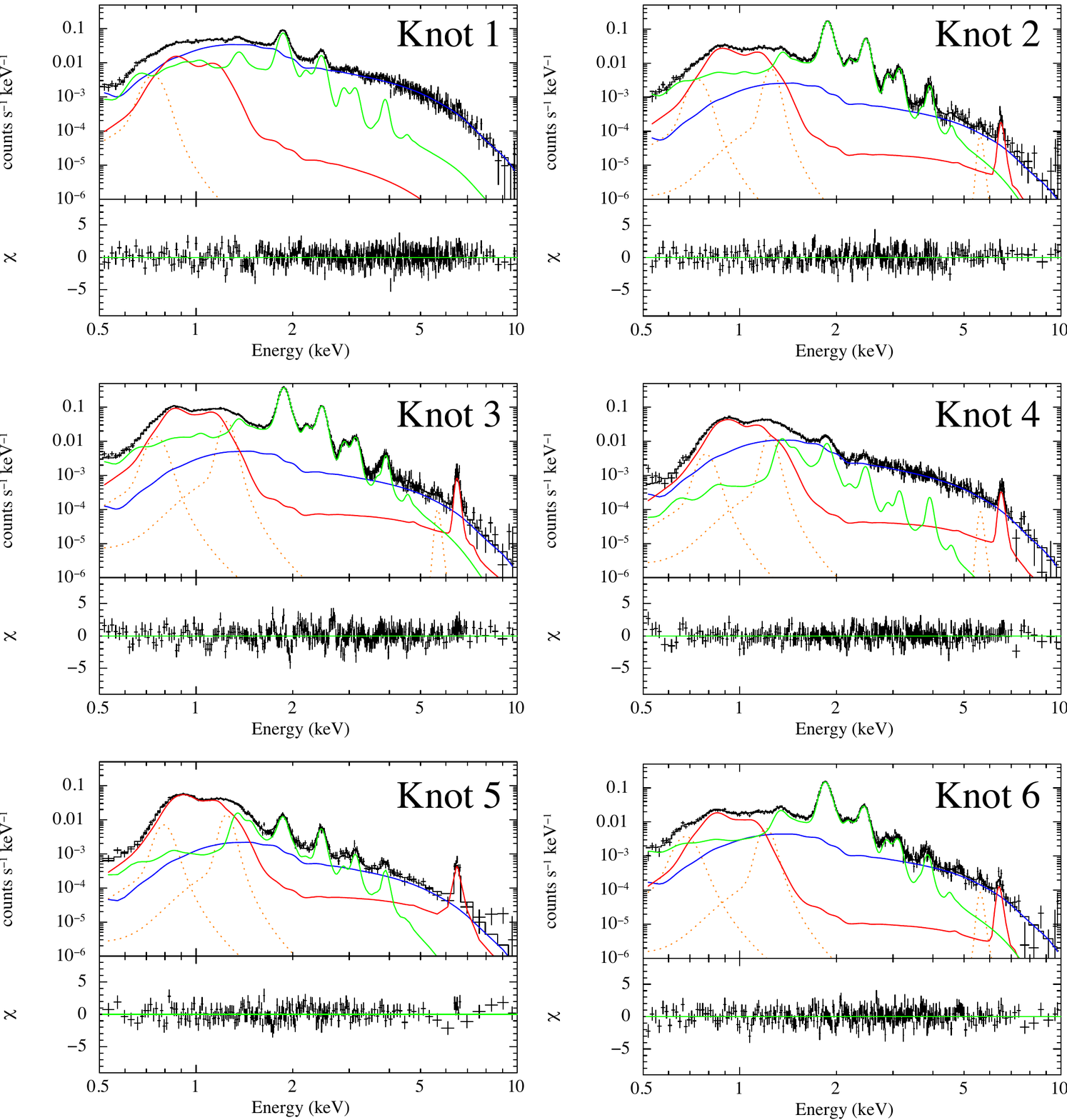}
 \end{center}
\caption{X-ray spectra and the best-fit models for the southeastern
  knots in \tycho. Labels in each panel (e.g., Knot1, Knot2, and so
  on) correspond to the region number in Figure \ref{fig:three_color}.
  Green and red curves show the model for the IME component and the
  iron component, respectively. Blue curves show the power-law
  continuum model. Orange dashed curves show the additional Gaussian
  models. Error bars on the spectra are shown at 1 $\sigma$}
\label{fig:SE_Knots}
\end{figure*}

\begin{table*}[t]
\tabletypesize{\small}
\caption{Fit Results for the Southeastern Knots}
\begin{center}
%\normalsize
\begin{tabular}{l|ccccccc}
\hline
Parameter 															&		Knot1						&		 	Knot2 					&		 	Knot3 	 				&		Knot4						&		 	Knot5 					&		 	Knot6 	 						\\ \hline
R.A. 																&	00$^{\rm h}$25$^{\rm m}$53$^{\rm s}$.647	&	00$^{\rm h}$25$^{\rm m}$57$^{\rm s}$.141	&	00$^{\rm h}$25$^{\rm m}$56$^{\rm s}$.998	&	00$^{\rm h}$25$^{\rm m}$57$^{\rm s}$.509	&	00$^{\rm h}$25$^{\rm m}$57$^{\rm s}$.627	&	00$^{\rm h}$25$^{\rm m}$51$^{\rm s}$.481			\\
Decl.	 															&	64$^\circ$08$^\prime$10$^{\prime\prime}$.69	&	64$^\circ$07$^\prime$47$^{\prime\prime}$.42	&	64$^\circ$07$^\prime$32$^{\prime\prime}$.52	&	64$^\circ$07$^\prime$05$^{\prime\prime}$.03	&	64$^\circ$06$^\prime$45$^{\prime\prime}$.35	&	64$^\circ$05$^\prime$42$^{\prime\prime}$.36			\\ \hline
$\chi^2$/d.o.f 														&		433/379					&		302/254					&		476/305					&		248/286					&		198/163					&			307/265						\\
$N_{\rm H}$ (10$^{22}$cm$^{-2}$) 									&	0.55$\pm$0.05					&	0.59$^{+0.03}_{-0.02}$			&	0.61$\pm0.01$		&	0.65$^{+0.05}_{-0.01}$		&	0.66$^{+0.01}_{-0.04}$			&		0.48$^{+0.03}_{-0.01}$				\\
Line broadening (eV)												&	29$\pm$2						&	23$\pm$1						&	25$\pm$1					&	16$^{+6}_{-9}$					&	15$^{+5}_{-6}$					&		24$\pm$2							\\
Velocity\tablenotemark{c} (km s$^{-1}$) 												&	$-$2330$^{+270}_{-10}$			&	$-$2410$^{+10}_{-100}$			&	$-$2393$^{+2}_{-10}$				&	$-$1880$^{+180}_{-100}$			&	$-$1050$^{+90}_{-160}$			&	$+$900$^{+50}_{-110}$						\\
{\underline {\bf IME component}} 									&									&		 		 					&		 				 	 		&									&		 		 					&		 				 	 				\\
$kT_{\rm e}$ (keV)													& 	2.6$^{+2.1}_{-1.0}$			& 	1.3$\pm0.1$			& 	1.23$\pm$0.04					& 	0.75$\pm$0.03					& 	0.60$\pm$0.01					&	1.7$^{+0.3}_{-0.2}$					\\
$n_{\rm e}t$ (10$^{10}$cm$^{-3}$ s)									& 	2.0$^{+0.5}_{-0.3}$				& 	4.8$^{+0.7}_{-0.2}$				& 	5.1$\pm$0.2					& 	28$^{+41}_{-5}$					& 	1000\tablenotemark{a}			&	3.1$\pm$0.6								\\
${\rm [Mg/C]/[Mg/C]}_{\odot}$														& 	1.4$^{+0.6}_{-0.3}$				&  	5.8$\pm$0.9						& 	2.8$\pm$0.1					& 	360$\pm30$				& 	110$\pm$6						&	32$^{+36}_{-2}$							\\
${\rm [Si/C]/[Si/C]}_{\odot}$																	& 	9$^{+4}_{-2}$				&	120$^{+40}_{-20}$				& 	61$\pm$10						& 	430$^{+160}_{-40}$				& 	140$^{+9}_{-7}$					&	430$^{+490}_{-10}$						\\
${\rm [S/C]/[S/C]}_{\odot}$															& 	10$^{+5}_{-3}$				&  	160$^{+40}_{-20}$				& 	80$\pm$13						& 	480$^{+100}_{-110}$				& 	400${\pm30}$				&	430$^{+20}_{-160}$						\\
${\rm [Ar/C]/[Ar/C]}_{\odot}$															& 	7$^{+5}_{-3}$				& 	180$^{+80}_{-30}$				& 	79$\pm$5						& 	870$^{+2900}_{-720}$			& 	780$^{+190}_{-180}$				&	320$\pm50$						\\
${\rm [Ca/C]/[Ca/C]}_{\odot}$															& 	22$^{+20}_{-14}$				& 	400$^{+180}_{-70}$				& 	180$\pm$20						& 	4000$^{+6000}_{-3000}$			& 	2000$^{+8000}_{-1000}$			&	900$^{+700}_{-200}$						\\
norm $\int n_e n_{\rm C} dV/4\pi d^2 /[{\rm C}/{\rm H}]_\odot$ ($10^{9}$ cm$^{-5}$)					& 	6$^{+4}_{-2}$				& 	1.9$^{+1.1}_{-0.7}$				& 	8.8$^{+0.1}_{-0.6}$			& 	0.057$^{+0.61}_{-0.001}$		& 	0.39$^{+0.57}_{-0.01}$		&	0.36$^{+0.23}_{-0.01}$					\\
{\underline {\bf Fe component}}										&									&									&									&									&		 		 					&		 				 	 				\\
$kT_{\rm e}$ (keV)													& 	1.2$^{+2.0}_{-0.5}$			& 	8.6$^{+0.5}_{-1.4}$			& 	9.3$\pm$0.1					& 	9.2$^{+0.4}_{-0.2}$			& 	9.1$\pm0.1$			&	10$^{+7}_{-1}$							\\
$n_{\rm e}t$ (10$^{10}$cm$^{-3}$ s)									& 	2$^{+3}_{-1}$			& 	1.31$\pm$0.05					& 	1.09$\pm$0.03					& 	1.58$\pm$0.03					& 	1.52$^{+0.04}_{-0.02}$			&	0.97$\pm$0.08							\\
${\rm [Fe/C]/[Fe/C]}_{\odot}$													& 	1.1$^{+0.9}_{-0.4}$			& 	5.8$^{+1.0}_{-0.9}$				& 	4.8$^{+0.9}_{-0.1}$			& 	360$^{+9600}_{-10}$				& 	70$^{+110}_{-2}$				&	18.4$^{+4.8}_{-0.7}$						\\
norm $\int n_e n_{\rm C} dV/4\pi d^2 / [{\rm C}/{\rm H}]_\odot$ ($10^{9}$ cm$^{-5}$)	& 	6\tablenotemark{b}							& 	1.9\tablenotemark{b}							& 	8.8\tablenotemark{b}							& 	0.057\tablenotemark{b}							& 	0.39\tablenotemark{b}							&	0.36\tablenotemark{b}									\\
{\underline {\bf power-law component}}								&									&									&									& 									& 									&											\\
$\Gamma$															& 	2.71$^{+0.06}_{-0.05}$			& 	2.4$\pm0.3$			& 	2.45$^{+0.07}_{-0.06}$			& 	2.83$\pm$0.03					& 	2.59$^{+0.09}_{-0.06}$			&	2.3$^{+0.2}_{-0.1}$					\\
norm  ($\times10^{-5}$ ph keV$^{-1}$ cm$^{-2}$ s$^{-1}$ at 1 keV)	& 	32$\pm2$			& 	2.3$^{+1.1}_{-0.7}$				& 	4.6$\pm$0.5					& 	11.80$^{+0.14}_{-0.03}$			& 	2.2$\pm0.2$			&	3.6$\pm$0.6								\\
{\underline {\bf Additional lines}}									&									&									&									&									&		 		 					&		 				 	 				\\
Fe L + O K Center (keV)												& 	0.73$^{+0.02}_{-0.03}$			& 	0.74$^{+0.02}_{-0.01}$		& 	0.75$\pm$0.01					& 	0.78$\pm$0.01					&  	0.80$\pm$0.01					&	0.68$^{+0.02}_{-0.01}$					\\
Fe L + O K norm	($\times10^{-6}$ ph cm$^{-2}$ s$^{-1}$)				& 	5$\pm2$			& 	3.5$^{+0.9}_{-1.1}$				& 	13.9$\pm$1.6					& 	3.3$^{+0.7}_{-0.8}$			& 	5.5$^{+0.7}_{-1.2}$			&	6.0$^{+1.4}_{-1.6}$						\\
Fe L Center (keV)											& 	---								& 	1.25$\pm$0.01					& 	1.247$\pm$0.004					& 	1.242$\pm$0.006					& 	1.247$\pm$0.006					&	1.21$^{+0.03}_{-0.01}$					\\
Fe L norm ($\times10^{-6}$ ph cm$^{-2}$ s$^{-1}$)			& 	---								& 	1.9$^{+0.5}_{-0.3}$			& 	11.2$\pm$0.7					& 	4.1$^{+0.7}_{-1.3}$			& 	4.0$\pm0.4$			&	2.5$^{+1.0}_{-1.1}$					\\
Cr K Center (keV)													& 	---								& 	5.6$^{+0.6}_{-0.1}$			& 	5.65$^{+0.14}_{-0.21}$			& 	5.61$\pm$0.11					& 	---								&	5.61$^{+0.62}_{-0.10}$					\\
Cr K norm	($\times10^{-8}$ ph cm$^{-2}$ s$^{-1}$)					& 	---								& 	7$\pm$6					& 	9$\pm$8					& 	8$\pm$8					& 	---								&	10$\pm9$					\\ \hline
\end{tabular}
\label{tab:SE_Knots}
\end{center}
\tablenotetext{1}{Value fixed to the equilibrium ionization limit. }
\tablenotetext{2}{Emission measure of Fe model component linked to the value for the IME component. }
\tablenotetext{3}{Velocities are based on the vnei model ``redshift'' parameter.}
\end{table*}

We extracted the spectra from these circular regions (with
radius 0.1$^\prime$) and fit them
  over the 0.5--10 keV energy band
with a two component NEI model, power-law continuum, and
additional Gaussian lines. The two NEI models account for iron
emission separately from the intermediate-mass elements (IMEs), Si, S,
Ar, and Ca. We link the redshift parameters for the two NEI
components. We assume no hydrogen, helium, or nitrogen in the shocked
SN Ia ejecta and quote abundances with respect to carbon (the
  lowest atomic number species that we include in our spectral
  analysis).  Oxygen and neon are kept fixed at their solar value with
  respect to carbon; the abundances of the other species are allowed
  to be free.  Several Gaussian lines at $\sim$0.7 keV, $\sim$1.2 keV
and $\sim$5.6 keV were added to account for missing lines, such as
 Fe-L \citep[$n = 3s, 3d \rightarrow 2p$ for Fe XVIII as shown
    in][]{2007ApJ...670.1504G} and/or O-K \citep[K-shell transition
    lines higher than K$\delta$ as shown in][]{2008PASJ...60S.141Y},
  Fe-L \citep[$n = 6,7,8 \rightarrow 2$ for Fe XVII, $n = 6,7
    \rightarrow 2$ for Fe XVIII, and $n = 6 \rightarrow 2$ for Fe XIX
    as shown in][]{2000ApJ...530..387B,2001A&A...365L.329A}, and
Cr-K$\alpha$, respectively, in the atomic databases. Line broadening
of the plasma models is included using the gsmooth model, and the
broadening of the additional Gaussian lines is linked to the same
value.  Absorption is included assuming solar abundances 
  \citep{1989GeCoA..53..197A}.

The spectral data and best-fit models are shown in Figure
\ref{fig:SE_Knots} and numerical values of the fit parameters are
given in Table \ref{tab:SE_Knots}.  The quoted abundances are
  relative to carbon relative to the solar value and in nearly all
  cases, as expected for SN Ia ejecta, are much greater than unity.
This model provided good results (with $\chi^2$/d.o.f $<$ 1.6), and
has highlighted some differences among the knots. Knot2, Knot3, and
Knot6 are more Si-rich, while Knot4 and Knot5 are more Fe-rich; both
points are consistent with previous results. Knot1 is dominated by
nonthermal emission compared to both the Si and Fe thermal components.
The velocities of the Si-rich and Fe-rich knots are $\sim-$2400 km
s$^{-1}$ and $\sim-$1900 to $-$1100 km s$^{-1}$, respectively. This is
unexpected for a spherical expansion, where the edge of the shell
should have no velocity along the line of sight.  These results
suggest that the SE knots are in fact inclined to the plane of the sky
and are therefore moving faster than their proper motion would
imply. We return to this point below.  Of course the reader should
keep in mind the systematic uncertainty of $\sim$500--2000 \kms\ in
the velocities of individual blobs as demonstrated in section
\ref{sec:blobs}.

%%%%%%%%%%%%%%%
%   Section 4
%%%%%%%%%%%%%%%

\section{Discussion}

Thanks to the high angular resolution of {\it Chandra}, we have
obtained (1) a more detailed view of the radial profiles of line
centroid and width, (2) consistency of our expansion velocity
measurements with previous results and (3) clear identification of
red- and blue-shifted components on multiple angular scales in \tycho.
These basic results hold the hope of advancing our understanding of
the type Ia SNe mechanism. In this section, we begin this effort as we
consider the implications of our results for the distance to \tycho,
the origin and nature of the southeastern (SE) knots, and the shock heating
processes in the ejecta.

\subsection{Distance to Tycho's SNR}

The expansion rate of \tycho\ has been studied using proper motion
measurements from X-ray imaging.  \cite{2010ApJ...709.1387K}
investigated the expansion rates of both the forward-shock and the
reverse-shocked ejecta using {\it Chandra} ACIS high-resolution images
of \tycho\ obtained in multiple epochs.  For the reverse-shocked
ejecta, they presented proper motion measurements for five azimuthal
sectors around the rim for two sets of epoch pairs \citep*[see Table 3
  in][]{2010ApJ...709.1387K}.  We use the 2003--2007 comparison (since
it uses the same instrument ACIS-I for both epochs) and average their
five azimuthal results to arrive at a mean proper motion of $\mu =
(0.267 \pm 0.056)^{\prime\prime}$ yr$^{-1}$. The uncertainty here is
taken to be the standard deviation of the five azimuthal values
(multiplied by 1.6 to approximate the 90\% confidence level), rather
than the uncertainty on the mean.  Combining this with our expansion
velocity of 5010$\pm$340 km s$^{-1}$, we estimate an allowed range on
the distance to \tycho\ of  $D = (4.0 \pm 0.3
  ~^{+1.0}_{-0.7})(V/5010 {\rm ~km~s}^{-1}) (\mu/0.267^{\prime\prime}
  {\rm ~yr}^{-1} )$ kpc, where the first and second terms show the
uncertainties from expansion velocity and proper motion.  This is
consistent with the result from {\it Suzaku}
\citep*[][]{2010ApJ...725..894H}: 4 $\pm$ 1 kpc as well as the result
based on the SN peak luminosity, as established by the observed
optical light-echo spectrum, and the maximum apparent brightness from
the historical records: 3.8$^{+1.5}_{-1.1}$ kpc
\citep*[][]{2008Natur.456..617K}.

Although the \chandra\ expansion speed measurement seems redundant in
that it reproduces the apparently more precise value from {\it
  Suzaku}, it is important to note that the {\it Suzaku} result is
subject to a correction factor due to that telescope's large PSF that
is completely eliminated in the case of \chandra.  Additionally,
  the high angular resolution of \chandra\ has allowed us to assess
  whether there is an intensity-dependent bias in the measured
  velocity difference, which might arise if, for example, brighter
  blobs tended to move at higher or lower speeds than fainter ones.
  We find that this is not a concern.  The velocity difference between
  high and low surface brightness regions in the central region of
  Tycho's SNR is small (less than 10\%) and not statistically
  significant.

Estimating the remnant's distance from the individual blob analyses
will be more uncertain due to the several systematic effects on the
velocity measurements (see \S 3.4) and because of difficulty in
identifying an appropriate matched sample of knots with good proper
motion measurements.

\subsection{High Velocity Knots in the Southeastern Quadrant}
The SE quadrant is one of the most mysterious features in
Tycho's SNR, and it is not yet understood how such a prominent
structure could be made. An aspherical explosion is one of the
possibilities. Theorists have found a number of ways to produce
asymmetric SN Ia explosions, including, pre-explosion convection
\citep{2006ApJ...640..407K}, off-center ignition of the burning front
  \citep{maeda+10,ropke+07}, and gravitationally confined detonations
  \citep{plewa+04,jordan+08}.  On the observational front,
  \cite{2010Natur.466...82M} have argued for large scale explosion
  asymmetries to explain the diversity in the spectral evolution of SN
  Ia.  In addition to \tycho, the elemental composition in SN~1006
  inferred from {\it Suzaku} observations also appears to be
  asymmetric \cite{2013ApJ...771...56U}.

The light-echo spectrum of Tycho's SNR \cite{2008Natur.456..617K},
spectrum shows a high velocity feature (HVF) identified as the Ca II
triplet at a velocity of 20,000--24,000 \kms\ during the early
evolutionary phase of the SN that Tycho observed. Similar HVFs have
been found in many SNe
\citep*[e.g.,][]{2005ApJ...623L..37M,2014MNRAS.437..338C}, as a result
of asphericity in the explosion due to, for example, accretion from a
companion or an intrinsic effect of the explosion itself
\citep*[][]{2003ApJ...591.1110W,2003ApJ...593..788K,2006ApJ...645..470T}.

In section \ref{sec:VSK}, we found that the SE knots have blueshifted
spectra; we adopt mean radial velocity values of 2400 km s$^{-1}$ for
the Si-rich knots and 1500 km s$^{-1}$ for the Fe-rich knots (note
that we restrict our discussion here to Knot2 through Knot5).
\cite{2010ApJ...709.1387K} determined the proper motions of these
knots. For the Si- and Fe-rich knots, the proper motions are
0.219--0.231 arcsec yr$^{-1}$ and 0.279--0.293 arcsec yr$^{-1}$,
respectively. Assuming a distance of 4.0 kpc, we can estimate
the transverse velocities as 4160--4380 \kms\ (Si) and 
  5290--5560 \kms\ (Fe). Combining with the radial velocities, yields
3-dimensional space velocities of 4800--5000 \kms\ for the
Si-rich knots, which are comparable to the Si expansion speed of the
rest of the remnant, and and 5500--5760 \kms\ for the Fe-rich
knots, which are some 33\%--44\% higher than the expansion speed
for Fe.  Although large, these values are not outside the range of
velocities we see elsewhere in the remnant
(Figure~\ref{fig:V}).

\begin{figure}[h]
 \begin{center}
  \includegraphics[trim=0 0 0 170,clip,width=8cm]{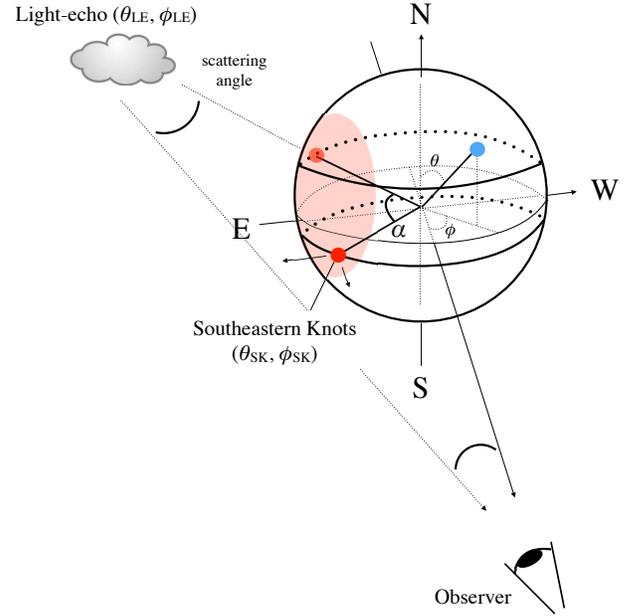}
 \end{center}
\caption{Schematic view of the positional relationship between the
  southeastern knots and the light-echo.}
\label{fig:LEK}
\end{figure}

Now we consider the relationship between the HVFs seen in the light
echo spectrum and the SE knots by examining the positional
relationship between them. Figure \ref{fig:LEK} shows a schematic view
to guide the discussion.  The plane of the sky lies in the plane
defined by the NS-EW axes and we indicate a sphere with unit radius
centered at the remnant's center.  We use a spherical polar coordinate
system ($\theta$,$\phi$) as defined by the blue dot on the unit
sphere.  We assume a distance of 4.0 kpc.

The SE knots (indicated by the labeled red dot on the unit sphere)
are slightly in front of the sky plane and are slightly south of the
EW axis.  According to \citet{2010ApJ...709.1387K} the
SE knots are located at angles of 97.5--107.5$^{\circ}$ from
north; we adopt the central value of $\theta_{\rm SK} =
102.5^{\circ}$.  We determine $\phi_{\rm SK}$ from the ratios of
transverse and radial velocities, namely 0.55--0.58 (Si-rich knots)
and 0.27--0.28 (Fe-rich knots).  We assume the mean ratio 
($\sim$0.42)
which yields a value of $\phi_{\rm SK} \sim 
 23^{\circ}+270^{\circ} =
293^{\circ}$.

The position on the sky of the light echo is about $3^{\circ}$ away
from the center of \tycho\ toward the NW and the polar angle is
$\theta_{\rm LE} = 62^{\circ}$; for our assumed distance of 4.0
  kpc to \tycho, the scattering angle is $65^{\circ}$. From these
values \citep[which we updated from the values in
][]{2008Natur.456..617K}, we determine that $\phi_{\rm LE} =
270^{\circ}-(90^{\circ}-3^{\circ}-65^{\circ}) = 248^{\circ}$.  Using
spherical trigonometry, we determine that there is an angular
separation of $\sim$$59^\circ$ between the centroid of the SE knots
and the viewing direction of the light echo.  The separation between
the Fe-rich knots and the light echo is about the same
$\sim$$59^\circ$.

There are systematic uncertainties on this result from the light
echo (whose location with respect to the remnant depends on the
assumed distance, since it must satisfy light-travel time arguments)
and uncertainty on the radial velocities (as discussed above) and
transverse velocities (whose main uncertainty is the assumed
distance).  Still we can set a robust lower limit on the angular
separation between these based on the very accurately determined
polar angles: $>$40$^\circ$ (for the mean of the SE knots) and
$>$45$^\circ$ (for the Fe-rich knots).

Three-dimensional models suggest that large blobs (opening angle:
$\sim$80$^{\circ}$) or a thick torus (opening angle:
$\sim$60$^{\circ}$) can naturally explain observations of the HVFs
\citep*[][]{2006ApJ...645..470T}.  Although the angular separation
between the knots and the direction to the light echo is similar to
the sizes of these proposed structures, the unique feature of the SE
quadrant is the presence there of Fe-rich knots, which are very
localized.  We therefore conclude that it is unlikely for the Fe-rich
knots in the SE quadrant to be responsible for the HVF in the light
echo spectrum.

Examining all six knots in the SE region (and ignoring systematic
uncertainties in radial and transverse velocities) we find that they
cover a full angular spread of $\Delta\theta\sim35^\circ$ and
$\Delta\phi \sim 40^\circ$. Yet how the knots are located is not
random; there is a correlation between $\theta$ and $\phi$. Near the
top of the feature (e.g., Knot1) $\phi \sim 299^\circ$, in the middle
(e.g., Knot4) $\phi \sim 289^\circ$, and at the bottom (e.g., Knot6)
$\phi \sim 260^\circ$.  The knots appear to be distributed in a chain
along the edge of the remnant and therefore form a distinct, fairly
compact, and kinematically connected structure in \tycho.

\begin{figure}[h]
 \begin{center}
  \includegraphics[trim=0 10 0 0,clip,width=8.5cm]{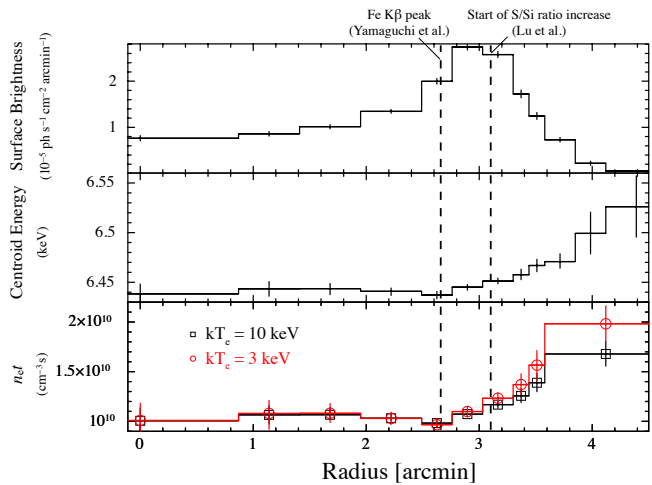}
 \end{center}
\caption{Radial profiles of the Fe-K$\alpha$ surface brightness (top),
  centroid energy (middle), and ionization age (bottom) for \tycho.
  The bottom panel is the result of spectral analysis using a
    vnei plus srcut model assuming fixed temperatures of 3 keV (red)
    and 10 keV (black) across the radial range shown. Dashed lines
  show the peak position of the Fe-K$\beta$ intensity
  \citep*{2014ApJ...780..136Y} and the location where the S/Si line
  ratio begins to increase while moving out from the remnant's center
  \citep*{2015ApJ...805..142L}.}
\label{fig:Fe}
\end{figure}

\subsection{Fe Ionization State Increase  at the Edge of \tycho}
\label{sec:fe-increase}
The ejecta are heated as the reverse shock propagates from the outside
of the remnant to the interior.  Thus the ionization age of the
shocked thermal plasma should vary with the time since the reverse
shock passed (ignoring variations in the density of the ejecta). This
is key information for our understanding of the heating processes at
the reverse shock. Some recent X-ray imaging and spectroscopy studies
have begun to shed light on this process.  One example is the work by
\cite{2014ApJ...780..136Y} on the variation of the Fe ionization state
near the reverse shock mentioned in the last section.  In other work,
\cite{2015ApJ...805..142L} found a systematic increase in the sulfur
to silicon K$\alpha$ line flux ratio with radius through the outer
edge of Tycho's SNR, which they interpreted as a radial dependence of
the ionization age.

In section \ref{sec:RP}, we found a strong increase in the Fe-K
centroid energy also at the outer edge of Tycho's SNR (see Figure
\ref{fig:Fe}).  The line centroid energy (middle panel) increases by
$\sim$90 eV over a radial distance of approximately
1$^\prime$. Carrying along in the same vein as the studies mentioned
in the previous paragraph, we interpret this change as being due to a
difference in the Fe ionization state and carried out spectral fits of
the Fe-K$\alpha$ band spectra using the srcut (continuum) and vnei
(thermal) models in XSPEC.  The temperature of the vnei model was
fixed at values of 3 keV and 10 keV; we used the ionization age
parameter ($n_e t$) to account for the observed changes in centroid
energy.  As before for the srcut model, we fixed the radio spectral
index to $\alpha = -0.65$.  Note that the scenario we investigate
  here is intended to be illustrative.  A future, more definitive
  study would consider the time evolution of temperature and density
  for a realistic ejecta density profile.

We found a gradual, modest increase of the ionization age from
$10^{10}$ cm$^{-3}$ s to $1.7\times 10^{10}$ cm$^{-3}$ s (10 keV) or
$2.0\times 10^{10}$ cm$^{-3}$ s (3 keV) over radii of
$\sim$2.8$^{\prime}$ to $\sim$3.8$^{\prime}$.  The inner radius is
close to the peak position of the Fe-K$\beta$ emission and also to
where the S/Si line ratio begins to increase moving out.  The
difference in Fe ionization age over the outer edge of \tycho\ is
$\Delta n_e t \sim (0.7-1.0)\times10^{10}\, {\rm cm^{-3} s} \sim
(220-320) (n_e/{\rm 1\, cm^{-3}})$ yr.  This is a plausible range
given the known age of \tycho\ (440 yr); additionally the ionization
timescale profile from the one-dimensional models of
\tycho\ \citep{2006ApJ...645.1373B} show a strong radial gradient
reaching values of $n_et \sim 2\times10^{10}\, {\rm cm^{-3} s}$ at the
edge of the ejecta, while the electron temperature remains
  relatively flat at a value of $\sim$3 keV.  However the radial
region over which we see this variation is exactly where 1D models
fail, i.e., where the Rayleigh-Taylor instability dominates the
structure of the remnant.  Understanding the thermodynamic evolution
of the plasma in this region will allow us to better understand and
model this important region.

%%%%%%%%%%%%%%%
%   Section 5
%%%%%%%%%%%%%%%

\section{Conclusions}

In this paper we have carried out a detailed analysis of the deep
\chandra\ ACIS-I observation ($\sim$734 ks in total exposure) of
\tycho.  We have presented measurements of ejecta velocities and have
investigated the heating processes in the ejecta.  Our results can be
summarized as follows.

\begin{enumerate}
\setlength{\parskip}{0.0cm}
  \setlength{\itemsep}{0.1cm}

 \item We investigated the radial dependence of the Si-He$\alpha$,
   S-He$\alpha$, and Fe K$\alpha$ line intensity, line centroid, and
   line width and obtained results consistent with previous work.
   \chandra's exceptional angular resolution allowed us to discover
   several new features in the radial profiles, including radial
   energy centroid shifts, a deep minimum in the line width profile
   for all species at a radius of $\sim$3.4$^\prime$, and a gradual
   increase in the Fe line centroid beyond radii of $\sim$3$^\prime$.
   From the line width profile we determined the expansion velocity of
   the Si-rich ejecta shell to be {$5010\pm340$ \kms} and, with the
   published proper motion of the Si-shell, obtained a distance
   measurement to Tycho's SNR of {$4.0 \pm 0.3 ~^{+1.0}_{-0.7}$} kpc.
   Although this is fully consistent with the previous {\it Suzaku}
   result, it is subject to fewer systematic uncertainties.

 \item The Si-He$\alpha$ line from \tycho\ shows large ($\sim$60 eV)
   energy centroid shifts across the image with the largest range
   appearing near the center of the remnant. The distribution of
   centroid shifts is structured on scales ranging from arcseconds to
   arcminutes, which agrees qualitatively with the highly structured
   intensity distribution.  Structure in the energy centroid image can
   be matched to features in the line centroid radial profile. We
   argue that these structures are due largely to differences in the
   intrinsic intensity of the approaching and receding hemispheres of
   the SNR.

\item We perform detailed spectral fits on 27 blobs using
  nonequiliubrium ionization thermal plasma models.  We succeed in
  separating these features cleanly into redshifted, blueshifted, and
  low velocity clumps of ejecta. The determination of velocities is
  shown to be robust with respect to other spectral fit parameters
  that can influence line centroids, such as the ionization age
  parameter. For a subset of the most rapidly moving blobs we perform
  joint fits with the ACIS-S data in order to establish the level of
  systematic velocity uncertainty: $\sim$500--2,000 \kms\ where the
  ACIS-S spectra tend to be more redshifted than ACIS-I.  Using a
  local background for each blob tends to increase fitted velocities
  by approximately 1000--2000 \kms. We conclude after considering
  these factors that the velocities of the redshifted and blueshifted
  blobs are $\lesssim 7,800$ km s$^{-1}$ and $\lesssim 5,000$ km
  s$^{-1}$, respectively.

\item
 We conclude based on geometric considerations that the unusual
 Fe-rich knots in the southeastern quadrant are not likely to be
 responsible for the high velocity Ca II absorption features seen in
 the light echo spectrum.  And if this exceptional set of knots is not
 responsible, then perhaps the origin of the HVF may be due to one of
 the more numerous, compact Si-rich knots that lie closer to the light
 echo direction.  Future spectral and kinematic studies of the knots
 in this direction may yield important clues to the nature of the HVF
 in SN Ia spectra.  A major step forward would be to obtain a light
 echo spectrum from a region off toward the SE of \tycho\ \citep[such
   as the fields 4523, 4821, and 5717 in][]{rest+08} that may provide
 a more direct view of the SE Fe-rich knots during the explosion.

\item We also note the detection of Cr K$\alpha$ line emission from 4
  out of the 6 SE knots analyzed.  Cr is also detected in most of the
  spectra from the radial profiles (at least out to Sky9). A careful
  study of the relative abundances of the Fe-group elements in the SE
  Fe-rich knots versus the rest of the remnant should in principle
  yield information about the explosion Si-burning processes in both
  regions.  

\item Finally, we found that the Fe-K$\alpha$ energy centroid showed a
  gradual increase beyond the radius of the peak intensity. We
  interpreted this as a difference in the elapsed ionization time by
  an amount $\Delta n_et \sim (220 - 320) (n_e/{\rm 1\, cm^{-3}})$
  cm$^{-3}$ yr since the material was shock heated. The region over
  which we see this happening is the region where Rayleigh-Taylor
  fingers of ejecta extend out into the forward shock region.
  Studying this region in more detail will yield useful information on
  this process.

\end{enumerate}

This work was initiated in preparation for observations of
\tycho\ with the {\it Hitomi} ({\it ASTRO-H}) satellite, which was
sadly lost in March 2016.  Our work with the \chandra\ ACIS detectors
shows the richness of the science that can be extracted from the
kinematics of SNRs.  There is still much that \chandra\ can do in this
area.  The \chandra\ High Energy Transmission Gratings have been used
to extract useful information on the kinematics of compact features in
the extended remnants Cas A \citep{lazendic+06} and G292.0+1.8
\citep{bhalerao+15} and an observation of Kepler's SNR (PI: Sangwook
Park) with a similar goal was observed in 2016 July.  Hopefully more
remnants will be observed in coming cycles.

\acknowledgments{ The authors are grateful to the {\it ASTRO-H}
  science team for giving us the opportunity to begin this
  collaboration during a visit by T.S.\ to Rutgers University in
  June--July 2015. We are also grateful for the travel support from
  Tokyo Metropolitan University for T.S.'s subsequent trip to Rutgers
  in January--April 2016. T.S.\ was also supported by the Japan
  Society for the Promotion of Science (JSPS) KAKENHI Grant Number
  16J03448.  The research was also supported in part by NASA grant
  NNX15AK71G. We thank the anonymous referee for a report that helped
  to improve the article. }

\bibliographystyle{yahapj}
%\bibliography{}

\appendix

\subsection{Separate Radial Dependence of the Red- and
Blueshifted Components of the Expanding Shell}

In section \ref{sec:blobs}, we showed that compact red- and
blueshifted blobs could be identified thanks to {\it Chandra}'s high
angular resolution.  We also know that the radial profile of line
width shows a gradual decline from a high value at the center
\citep*[e.g.,][]{2009ApJ...693L..61F,2010ApJ...725..894H}, consistent
with the scenario that we are seeing the combination of two different
velocity components of an expanding shell.  We now describe an
additional analysis, that used {\it Chandra}'s high resolution in a
different way, to separate the two shell components and provide
additional confidence for the expanding shell scenario.

\begin{figure*}[h]
 \begin{center}
  \includegraphics[trim=0 0 0 200,clip,width=15cm]{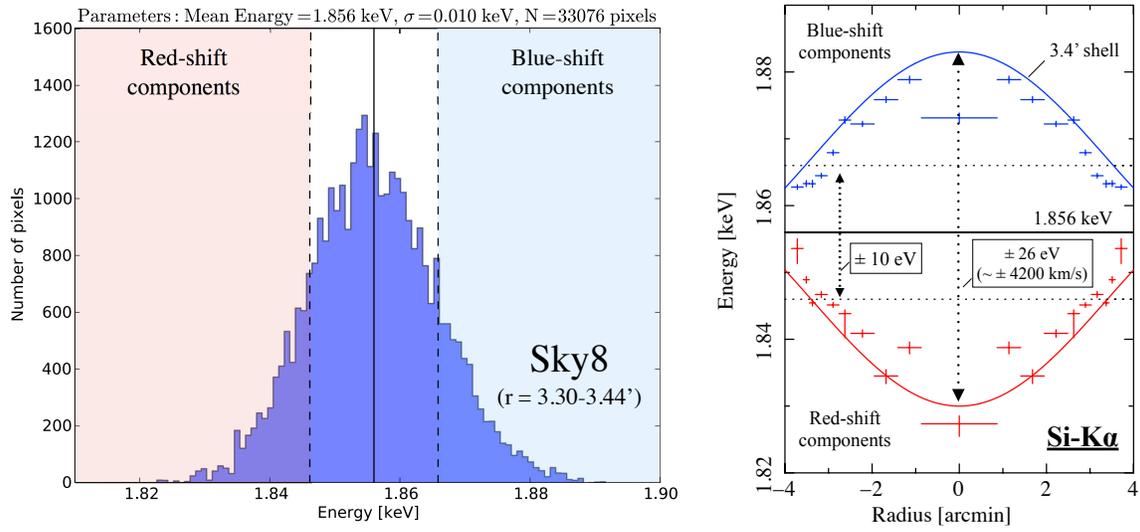}
 \end{center}
\caption{Left: the histogram of mean photon energies from the Sky8
  region (between radii of 3.30$^{\prime}$--3.44$^{\prime}$). Right:
  the radial dependence of the Si-He$\alpha$ centroid energies of the red- and
  blueshifted shell components. The blue and red solid lines show a
  cosine functions which approximatges the 3.4$^{\prime}$ shell
  expansion. This figure is symmetric for positive and negative
  radius.}
\label{fig:RE}
\end{figure*}

For each radial region we generated the distribution of Si-He$\alpha$
line centroid energies (Figure \ref{fig:RE}, left panel) from the
centroid map (Figure~\ref{fig:img}, left panel).  From this we
calculated the mean photon energy and standard deviation of the
distribution.  Two spectra were extracted for each radial region: one
for all pixels with energy centroids more than 1 standard deviation
above the mean (the blueshifted spectrum) and the other for all pixels
with centroids more than 1 standard deviation below the mean (the
redshifted spectrum).  Then we conducted spectral fitting (using the
same spectral model as in \S~\ref{sec:RP}) for each pair of spectra
from each region to obtain the centroid energies for each component.
Figure \ref{fig:RE} (right panel) shows the radial dependence of the
Si-He$\alpha$ centroid energy for these two components.  The general
shapes of these two curves is consistent with the projection of each
hemisphere (receding and approaching).

We approximate the projection effect with a pair of simple cosine
functions (Figure \ref{fig:RE} right).  The curves are not a fit to
the data.  We took the values of the mean energy and standard
deviation (1.856 keV and 10 eV) from the Sky8 region (the region
closest to the edge of the Si shell) and added to these the cosine
function for a shell of radus 3.4$^\prime$ and peak velocity of 4200
\kms.  The data points show a similar trend, but seem to prefer a
lower shell velocity.  This is to be expected since the red- and
blueshifted lines are broad (see Figure~\ref{fig:specdoublegauss}, top
panel) and so the higher velocity pixels are diluted by the more
numerous lower velocity ones.  We also see the effect of the shell
patchiness in the central radial bin. Our analysis here confirms that
this region includes much more redshifted than blueshifted emission.

\end{document}